\makeatletter
\def\input@path{{tex/}}
\makeatother
\documentclass[final,1p,times]{elsarticle}

\newcounter{bla}

\journal{Computer Physics Communications}

\usepackage{graphicx}
\usepackage[justification=RaggedRight]{caption}
\usepackage{subcaption}
\usepackage{bm}
\usepackage{amssymb}
\usepackage{amsmath}
\usepackage{amsfonts}
\usepackage[utf8]{inputenc}
\usepackage{placeins}
\usepackage{booktabs}
\usepackage{gensymb}
\usepackage[colorlinks=true,allcolors=blue]{hyperref}
\usepackage{microtype}
\usepackage{multirow, makecell}
\usepackage{lipsum}
\usepackage[capitalise]{cleveref}

\usepackage[utf8]{inputenc}

\usepackage{listings}
\usepackage{siunitx}
\usepackage{xspace}
\usepackage{comment}
\usepackage{floatrow}
\usepackage[T5,T1]{fontenc}
\usepackage{footmisc}

\usepackage{ragged2e}

\usepackage{placeins}

\usepackage{tikz}
\usetikzlibrary{trees}

\tikzstyle{every node}=[draw=black,thick,anchor=west]
\tikzstyle{selected}=[dashed,draw=red,fill=red!30]
\tikzstyle{optional}=[dashed,fill=gray!50]

\DeclareSIUnit \s {\second}
\DeclareSIUnit \ns {\nano\second}
\DeclareSIUnit \mus {\micro\second}
\DeclareSIUnit \ms {\milli\second}
\DeclareSIUnit \MB {\mega\byte}
\DeclareSIUnit \GB {\giga\byte}
\DeclareSIUnit \TB {\tera\byte}
\DeclareSIUnit \PB {\peta\byte}
\DeclareSIUnit \Mbps {\mega\bit/\s}
\DeclareSIUnit \Gbps {\giga\bit/\s}
\DeclareSIUnit \Tbps {\tera\bit/\s}
\DeclareSIUnit \Pbps {\peta\bit/\s}
\DeclareSIUnit \kton {\kilo\tonne} 
\DeclareSIUnit \kt {\kilo\tonne}
\DeclareSIUnit \Mt {\mega\tonne}
\DeclareSIUnit \eV {\electronvolt}
\DeclareSIUnit \keV {\kilo\electronvolt}
\DeclareSIUnit \MeV {\mega\electronvolt}
\DeclareSIUnit \GeV {\giga\electronvolt}
\DeclareSIUnit \PeV {\peta\electronvolt}
\DeclareSIUnit \EeV {\exa\electronvolt}
\DeclareSIUnit \ZeV {\zetta\electronvolt}
\DeclareSIUnit \m {\meter}
\DeclareSIUnit \cm {\centi\meter}
\DeclareSIUnit \in {\inchcommand}
\DeclareSIUnit \km {\kilo\meter}
\DeclareSIUnit \kV {\kilo\volt}
\DeclareSIUnit \kW {\kilo\watt}
\DeclareSIUnit \MW {\mega\watt}
\DeclareSIUnit \MHz {\mega\hertz}
\DeclareSIUnit \mrad {\milli\radian}
\DeclareSIUnit \year {years}
\DeclareSIUnit \POT {POT}
\DeclareSIUnit \sig {$\sigma$}
\DeclareSIUnit\parsec{pc}
\DeclareSIUnit\lightyear{ly}
\DeclareSIUnit\foot{ft}
\DeclareSIUnit\ft{ft}
\DeclareSIUnit \ppb{ppb}
\DeclareSIUnit \ppt{ppt}
\DeclareSIUnit \samples{S}
\DeclareSIUnit \pe{PE}
\DeclareSIUnit \mwe{mwe}

\newcommand{\enu}{\E_\enu}

\numberwithin{figure}{section}
\numberwithin{table}{section}
\numberwithin{equation}{section}

\lstdefinestyle{mystyle}{basicstyle=\ttfamily,
    breakatwhitespace=false,
    breaklines=true,
    captionpos=b,
    keepspaces=true,
    showspaces=false,
    showstringspaces=false,
    showtabs=false,
    tabsize=4
}
\AtBeginDocument{\heavyrulewidth=.08em
\lightrulewidth=.05em
\cmidrulewidth=.03em
\belowrulesep=.65ex
\belowbottomsep=0pt
\aboverulesep=.4ex
\abovetopsep=0pt
\cmidrulesep=\doublerulesep{}
\cmidrulekern=.5em
\defaultaddspace=.5em
}

\newcommand{\ttf}{\ttfamily}

\newcommand{\LeptonInjector}{\texttt{LeptonInjector}}
\newcommand{\LeptonWeighter}{\texttt{LeptonWeighter}}
\newcommand{\NuGen}{\texttt{NuGen}}
\newcommand{\ANIS}{\texttt{ANIS}}
\newcommand{\Python}{\texttt{Python }}

\newcommand{\PHOTOSPLINE}{\texttt{Photospline}}

\newcommand{\hdf}{\texttt{HDF5}}
\newcommand{\boostpython}{\texttt{boost-python}}
\newcommand{\nusim}{\texttt{NUSIM}}

\newcommand{\refeq}[1]{Eq.~(\ref{#1})}

\newcommand{\reftab}[1]{Table~\ref{#1}}

\DeclareFixedFont{\ttb}{T1}{txtt}{bx}{n}{12} 
\DeclareFixedFont{\ttm}{T1}{txtt}{m}{n}{12}  

\usepackage{color}
\definecolor{deepblue}{rgb}{0,0,0.5}
\definecolor{deepred}{rgb}{0.6,0,0}
\definecolor{deepgreen}{rgb}{0,0.5,0}

\usepackage{listings}

\newcommand\pythonstyle{\lstset{
language=Python,
basicstyle=\footnotesize,
otherkeywords={self},             
keywordstyle=\color{deepblue},
emph={MyClass,__init__},          
emphstyle=\color{deepred},    
stringstyle=\color{deepgreen},
showstringspaces=false            %
}}

\lstnewenvironment{python}[1][]
{
\pythonstyle
\lstset{#1}
}
{}

\newcommand\pythonexternal[2][]{{
\pythonstyle
\lstinputlisting[#1]{#2}}}

\newcommand\pythoninline[1]{{\pythonstyle\lstinline!#1!}}

\begin{document}
\begin{frontmatter}
\captionsetup[figure]{labelfont={bf},labelformat={default},labelsep={period},name={Figure},justification=RaggedRight}
\captionsetup[table]{labelfont={bf},labelformat={default},labelsep={period},name={Table},justification=RaggedRight}

\title{LeptonInjector and LeptonWeighter: A neutrino event generator and weighter for neutrino observatories}


\date{\today}

\author[loyola]{R. Abbasi}
\author[zeuthen]{M. Ackermann}
\author[christchurch]{J. Adams}
\author[brusselslibre]{J. A. Aguilar}
\author[copenhagen]{M. Ahlers}
\author[stockholmokc]{M. Ahrens}
\author[geneva]{C. Alispach}
\author[karlsruhe]{A. A. Alves Jr.}
\author[bartol]{N. M. Amin}
\author[harvard]{R. An}
% ...
\address[brusselslibre]{Universit{\'e} Libre de Bruxelles, Science Faculty CP230, B-1050 Brussels, Belgium}
% ...
\address[harvard]{Department of Physics and Laboratory for Particle Physics and Cosmology, Harvard University, Cambridge, MA 02138, USA}
% ...
\address[loyola]{Department of Physics, Loyola University Chicago, Chicago, IL 60660, USA}
\address[christchurch]{Dept. of Physics and Astronomy, University of Canterbury, Private Bag 4800, Christchurch, New Zealand}
% ...
\address[copenhagen]{Niels Bohr Institute, University of Copenhagen, DK-2100 Copenhagen, Denmark}
% ...
\address[geneva]{D{\'e}partement de physique nucl{\'e}aire et corpusculaire, Universit{\'e} de Gen{\`e}ve, CH-1211 Gen{\`e}ve, Switzerland}
% ...
\address[karlsruhe]{Karlsruhe Institute of Technology, Institute for Astroparticle Physics, D-76021 Karlsruhe, Germany }
% ...
\address[bartol]{Bartol Research Institute and Dept. of Physics and Astronomy, University of Delaware, Newark, DE 19716, USA}
% ...
\address[stockholmokc]{Oskar Klein Centre and Dept. of Physics, Stockholm University, SE-10691 Stockholm, Sweden}
% ...
\address[zeuthen]{DESY, D-15738 Zeuthen, Germany}
\fntext[padova]{also at Universit{\`a} di Padova, I-35131 Padova, Italy}
\fntext[mephi]{also at National Research Nuclear University, Moscow Engineering Physics Institute (MEPhI), Moscow 115409, Russia}
\fntext[tokyofn]{also at Earthquake Research Institute, University of Tokyo, Bunkyo, Tokyo 113-0032, Japan}

\begin{abstract}
We present a high-energy neutrino event generator, called \texttt{LeptonInjector}, alongside an event weighter, called \texttt{LeptonWeighter}.
Both are designed for large-volume Cherenkov neutrino telescopes such as IceCube.
The neutrino event generator allows for quick and flexible simulation of neutrino events within and around the detector volume, and implements the leading Standard Model neutrino interaction processes relevant for neutrino observatories: neutrino-nucleon deep-inelastic scattering and neutrino-electron annihilation.
In this paper, we discuss the event generation algorithm, the weighting algorithm, and the main functions of the publicly available code, with examples.
\end{abstract}

\end{frontmatter}

\noindent Program Summary: \\

\noindent\textit{Program Titles:} \texttt{LeptonInjector} and \texttt{LeptonWeighter} \\
\textit{CPC Library link to program files:} (to be added by Technical Editor)\\
\textit{Developer's repository links:} \url{https://github.com/icecube/LeptonInjector} and \url{https://github.com/icecube/LeptonWeighter} \\
\textit{Licensing provisions:} GNU Lesser General Public License, version 3. \\
\textit{Programming Language:} C++11 \\
\textit{External Routines: }  \\
\begin{itemize}
    \item Boost 
    \item HDF5
     \item nuflux (\url{https://github.com/icecube/nuflux})
      \item nuSQuIDS (\url{https://github.com/arguelles/nuSQuIDS})
    \item Photospline (\url{https://github.com/icecube/photospline})
    \item SuiteSparse (\url{https://github.com/DrTimothyAldenDavis/SuiteSparse})
\end{itemize}
\textit{Nature of problem:} \texttt{LeptonInjector}: Generate neutrino interaction events of all possible topologies and energies throughout and around a detector volume.

\texttt{LeptonWeighter}: Reweight Monte Carlo events, generated by a set of \texttt{LeptonInjector} Generators, to any desired physical neutrino flux or cross section.
 \\
\textit{Solution method:} \texttt{LeptonInjector}: Projected ranges of generated leptons and the extent of the detector, in terms of column depth, are used to inject events in and around the detector volume. Event kinematics follow distributions provided in cross section files.

\texttt{LeptonWeighter}: Event generation probabilities are calculated for each Generator, which are then combined into a generation weight and used to calculate an overall event weight. \\

\section{\label{sec:introduction} Introduction}

Neutrinos have been measured in a wide energy range from $\si\MeV$ energies in solar and reactor experiments to $\si\PeV$ energies in neutrino telescopes~\cite{Vitagliano:2019yzm}.
Different neutrino interaction processes~\cite{Formaggio:2013kya} are relevant in this wide energy range, from \textit{e.g.} coherent-neutrino scattering~\cite{Akimov1123} at very small momentum ($Q^2$) transfer, to very large $Q^2$ processes which create $W$ bosons~\cite{Glashow:1960zz,Seckel:1997kk,Alikhanov:2014uja,Zhou:2019vxt,Beacom:2019pzs} and heavy quark flavors~\cite{Barge:2016uzn}.
However, deep-inelastic scattering~\cite{Gandhi:1995tf} is always the dominant process above $\sim\SI{10}\GeV$.
This broad energy range has led to the development of various neutrino event generators used by experiments to simulate neutrino interactions~\cite{Hayato:2002sd,Andreopoulos:2009rq,Golan:2012rfa,Lalakulich:2011eh}, most of which have been optimized for $\si\GeV$ neutrino energy ranges and sub-megaton target mass detectors~\cite{Andreopoulos:2009rq}.
Such generators are not optimal for gigaton-scale neutrino detectors, often known as neutrino telescopes, such as the currently operating IceCube Neutrino Observatory at the Amundsen-Scott South Pole Station~\cite{Aartsen:2016nxy} and next-generation observatories such as KM3NeT~\cite{Adrian-Martinez:2016fdl} in the Mediterranean Sea and GVD in Lake Baikal~\cite{Avrorin:2018ijk}.

The first neutrino telescope event generators started their simulation at the Earth's surface~\cite{osti_5884484,Hill:1996hzh,Gazizov:2004va,Yoshida:2003js}, which required solving two distinct problems: neutrino transport through the planet and the generation of neutrino events near the sensitive volume.
The first such event generator was \nusim{}~\cite{Hill:1996hzh}, developed in the 1990s for the Antarctic Muon and Neutrino Detector Array (AMANDA); see also~\cite{Bailey:2002} for a similar effort for ANTARES. 
\nusim{} established the fundamental concepts of what would later evolve into this project by breaking the problem of event generation into a three-step procedure. 
First a neutrino energy was randomly drawn from a prior distribution, then forced to interact somewhere near the detector, and finally an event weight would be calculated and applied~\cite{Hill:1996hzh}.
This process relied on costly calculations of survival probability of the neutrino through the entirety of the Earth, and tightly coupled the generation and interaction of each neutrino to the calculation of its event weight.

\begin{figure}
    \centering
    \includegraphics[width=\linewidth]{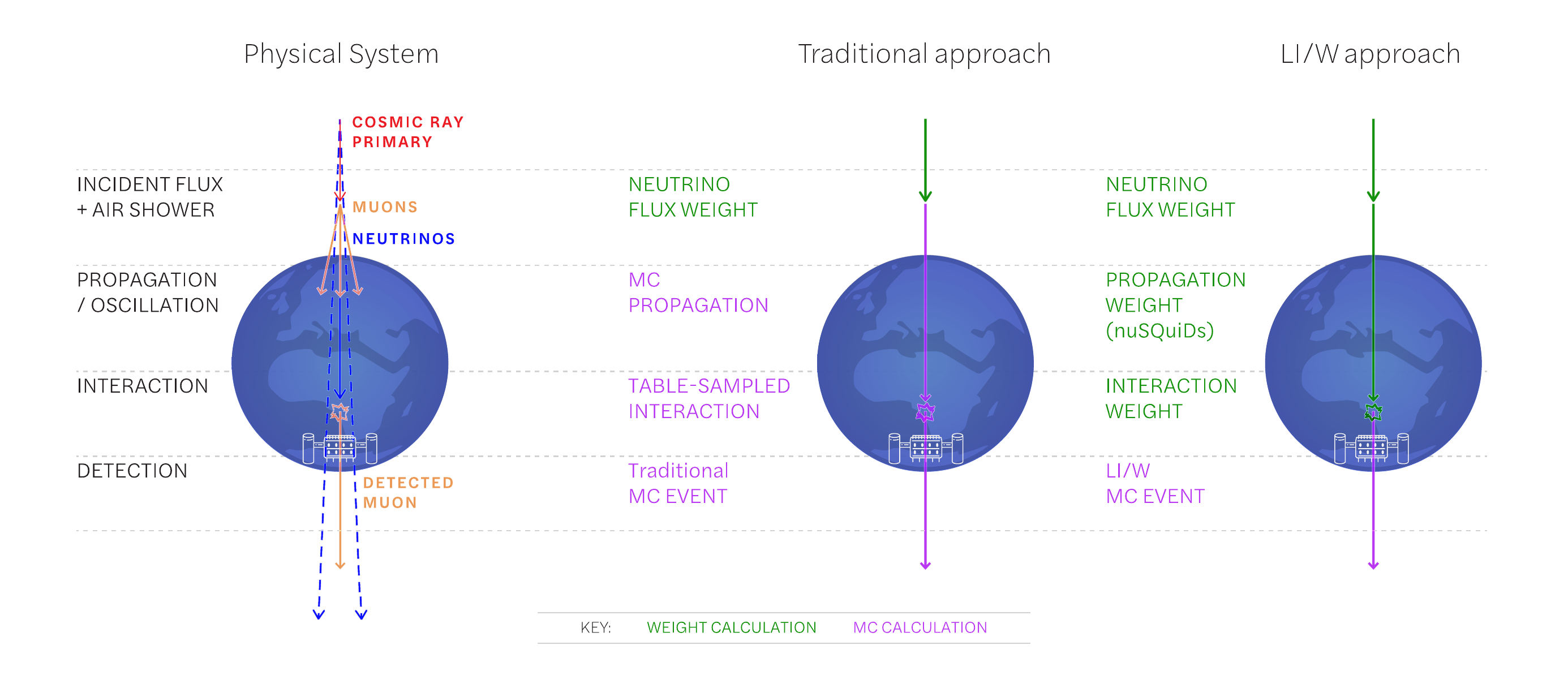}
    \caption{A diagram illustrating the different event generation and weighting steps for traditional methods compared with the \LeptonInjector{} and \LeptonWeighter{} philosophy.}
    \label{fig:nufsgen}
\end{figure}

In 2005, \nusim{} was ported to {\ttf C++} and released as the All Neutrino Interaction Simulation (\ANIS)~\cite{Gazizov:2004va}, and then modified and adopted into the IceCube internal framework~\cite{DeYoung:865626} as neutrino-generator, or \NuGen{}. 
The basic simulation scheme remained unchanged, although with each update of the software the scope of features grew and the fundamental features and techniques of the algorithm were further refined and optimized. 
In the past five years, efficient algorithms to solve the neutrino transport problem have become publicly available~\cite{Yoshida:2003js,Arguelles:2020hss,arguelles:2015nu,Zas:2017xdj,Vincent:2017svp,Safa:2019ege,Garcia:2020jwr}, allowing the possibility of simplifying event generation to only consider the problem of event generation in and around a volume near the detector.
This allowed the event generation scheme to be separated into two standalone and publicly-available software projects: \LeptonInjector{}~\cite{LeptonInjectorRepository} and \LeptonWeighter{}~\cite{LeptonWeighterRepository}.
This separation is not only convenient from software maintenance point of view, but also facilitates optimizations in different energy ranges. 
For example, IceCube's analyses focusing in $\si\EeV$ energies~\cite{Aartsen:2018vtx}, where the Earth is opaque to neutrinos, have used the JAVA-based JULIeT~\cite{Yoshida:2003js,shigeruyoshida_2020_4018117} software package for neutrino transport.
JULIeT, much like the {\ttf C++}-based nuSQuIDS~\cite{Arguelles:2020hss,arguelles:2015nu} package, has the computational advantage of solving Earth propagation using a set of differential equations instead of a Monte Carlo approach.
The combination of software presented in this work allows for the user to choose the neutrino transport solution that best suits their needs.
This simulation technique, the \LeptonInjector{}/\LeptonWeighter{} (LI/W), and traditional techniques are illustrated in Figure~\ref{fig:nufsgen}

In this paper, we will describe the structure and function of the LeptonInjector software package, as well as a companion package called LeptonWeighter.
In Section II we describe the basic functionality of LeptonInjector, focusing on the structure of the software (Section II.A), the injection of particles into the detector (Section II.B), and a comparison between the output of LeptonInjector and \NuGen{}.
Section III contains a description of LeptonWeighter and provides examples of reweighted neutrino samples from various physical sources of neutrinos.
We conclude in Section IV.
Details of the event and file structures provided by the software packages, as well as example driver scripts, are provided in Supplemental Material.

\section{\LeptonInjector\label{sec:overview}}

\LeptonInjector{} is written in {\ttf C++} with \boostpython{} bindings, and uses \PHOTOSPLINE~\cite{WHITEHORN20132214} for the cross sections needed for kinematic variable sampling.
A standalone version of the code is publicly available from the IceCube GitHub repository~\cite{LeptonInjectorRepository}.
In the description of the software that follows, we use \texttt{monospace} font to refer to libraries and packages, \textbf{bold} font to refer to classes, and \textit{italic} font to refer to members of a class..

\LeptonInjector{} is capable of simulating neutrino events of all flavors over a wide range of energies from $\SI{10}\GeV$ to $\SI{100}\PeV$ and beyond, undergoing neutrino-nucleon interaction in the Deep Inelastic Scattering (DIS) regime and antineutrino-electron scattering producing $W$ in a Glashow Resonance (GR) interaction, $(\bar{\nu}_{e}+e^{-} \to W^{-})$.
The initial event energy is sampled according to a single power-law spectrum at any desired spectral index, and final state kinematics are sampled from spline interpolations of the differential cross sections for the relevant interaction.
These splines are saved in \texttt{FITS} files generated by \PHOTOSPLINE~\cite{WHITEHORN20132214}.
For optimum efficiency, the spectrum of generated events would match the physical one.
Atmospheric and astrophysical neutrino fluxes, for example, follow a power-law flux.
As it is often desirable to maintain large sample size at high energies, events can be generated at one flux and subsequently reweighted to any physical flux using \LeptonInjector{}'s sister software package \LeptonWeighter{}, available at~\cite{LeptonWeighterRepository}.
Because the event generation starts from near the detector, the primary neutrino energy will be guaranteed to follow the spectral index of event generation; this is not the case for event generators beginning at the Earth's surface.

To facilitate the reweighting, \LeptonInjector{} creates configuration objects complete with a full description of all relevant event generation parameters.
\LeptonWeighter{} uses these configuration files to reweight events to any desired physical distribution. 

Following the event generation, described in this work, the IceCube Monte Carlo proceeds using the following publicly available packages.
First, leptons are propagated using {\ttf PROPOSAL}~\cite{Koehne:2013gpa}, a software package based on {\ttf MMC}~\cite{Chirkin:2004hz}, while hadronic or electromagnetic showers are propagated by the {\ttf Cascade Monte Carlo} (CMC) package, which implements the physics described in~\cite{Niess_2006} and~\cite{Wiebusch:thesis,2012APh....38...53R,Radel:2012ij}.
Next, Cherenkov photons arising from the charged particles are simulated by direct photon propagation using {\ttf CLSim}~\cite{9041727,CLSim} or {\ttf PPC}~\cite{Chirkin:2015kga,PPCStandAlone}.
Finally, a detector response is produced using a proprietary detector simulation.

\LeptonInjector{} requires two kinds of objects: one or more \textbf{Injectors} and a \textbf{Controller}.
An \textbf{Injector} object represents one primary neutrino type and one interaction channel, one cross section model to guide interactions, a number of neutrinos to be injected, and one parameter to control the sampling of the interaction vertex: ranged or volume mode, which are described in Section~\ref{sec:injection}. 
In practice, the type of primary neutrino and interaction channel are specified in the \textbf{Injector} as a pair of particles that would be generated in such an interaction. 
These final state particles are called \textit{finalType1} and \textit{finalType2}, and the order of these particles are strictly defined in Table~\ref{tbl:final_types}. 
A `Hadrons' particle is used to represent the hadronic shower produced by the recoiling nucleus from the DIS interaction and the hadronic decay channel from a $W$ produced in a GR interaction. 
Ideally the propagation of the hadronic showers would be simulated directly using GEANT4 or a similar particle interaction framework, although this process is far too computationally expensive to be practical. 
Instead, a parametrization is used for the propagation of the hadronic shower, developed through GEANT4 simulations, as described in~\cite{Wiebusch:thesis}.

\begin{table}[h]
\centering
    \begin{tabular}{c c c c}
	\toprule
	Event Type & Interaction & \textit{finalType1} & \textit{finalType2} \\
	\midrule
	Nu\{E,Mu,Tau\} & CC & \{E,Mu,Tau\}Minus & Hadrons \\
	Nu\{E,Mu,Tau\} & NC & Nu\{E,Mu,Tau\} & Hadrons \\
	Nu\{E,Mu,Tau\}Bar & CC & \{E,Mu,Tau\}Plus & Hadrons \\
	Nu\{E,Mu,Tau\}Bar & NC & Nu\{E,Mu,Tau\}Bar & Hadrons \\
	NuEBar & GR & Hadrons & Hadrons \\
	NuEBar & GR & \{E,Mu,Tau\}Minus & Nu\{E,Mu,Tau\}Bar \\
	\bottomrule
	\end{tabular}
    \caption{\textbf{\textit{Final State Particles.}}
    Final state particle types are given, in the right two columns, for various possible desired interactions.
    }
    \label{tbl:final_types}
\end{table}

\begin{figure}[p]
    \centering
    \makebox[\linewidth]{
        \includegraphics[width=1.5\linewidth]{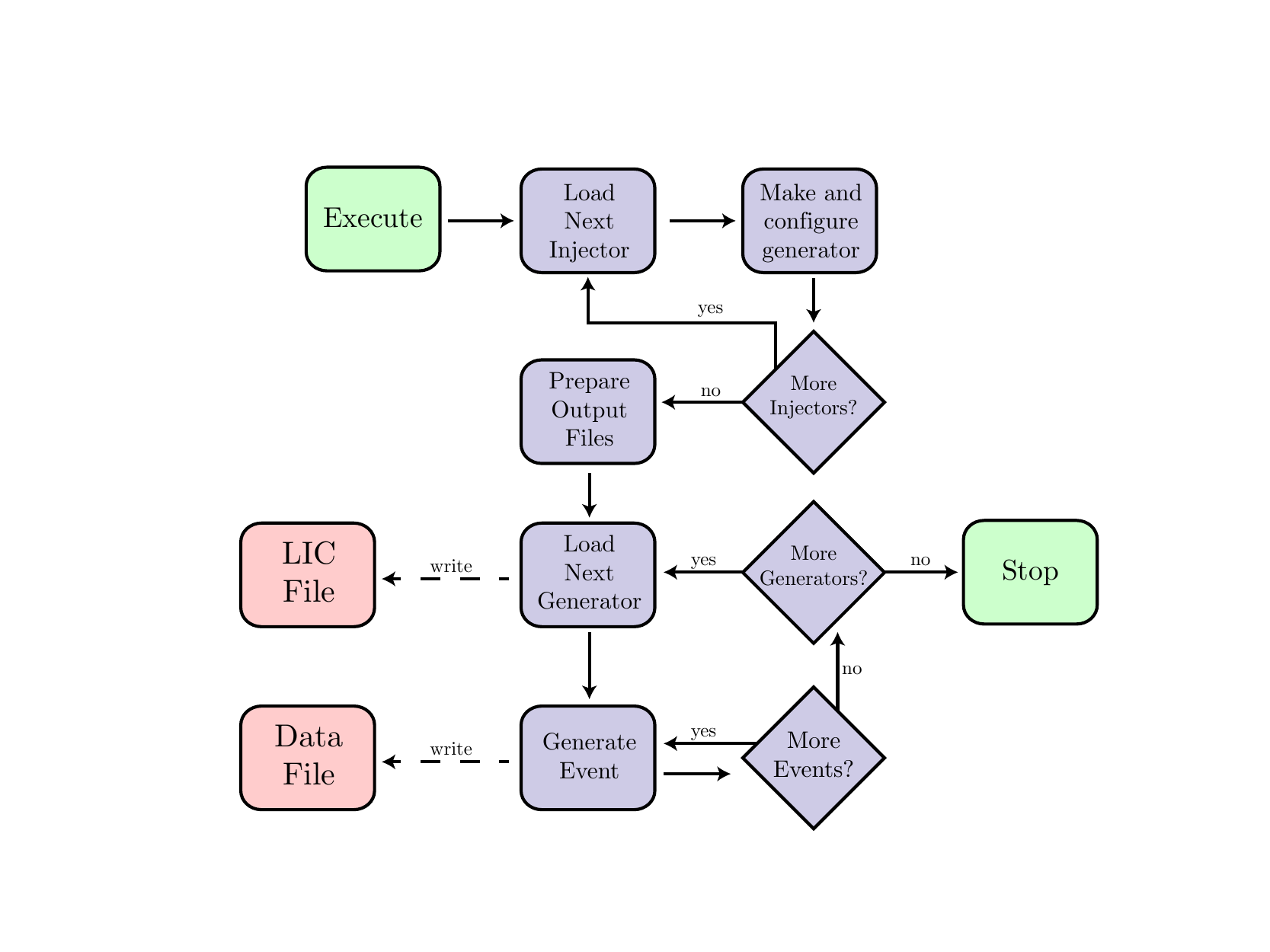}
    }
    \caption{Flowchart displaying the iterative process by a \textbf{Controller} prepares to generate, and then generates events. The \textbf{Generator} object is used by \LeptonInjector{} to store all necessary information to simulate events.}\label{fig:flow_LI}
\end{figure}

The \textbf{Controller} defines energy ranges, azimuth and zenith ranges, and a spectral index of the primary neutrino as shown in Table~\ref{tbl:injection_parameters}. 
One or more \textbf{Injector} objects must be assigned to a Controller as well as the destinations for the output files. 
Once the \textbf{Controller} is configured, the simulation is initialized by calling the \textit{Execute} member function.
The Controller iterates over its member \textbf{Injectors}, with each combining the \textbf{Controller}'s flux properties with its own injection parameters into a \textbf{Generator}, and generating events until reaching their target number as set by the user. 
This process is illustrated in Figure~\ref{fig:flow_LI}.
In both ranged and volume injection modes, \textbf{Injectors} generate a `primary' neutrino according to a power-law spectrum with the configured spectral index. 
This `primary' neutrino is the neutrino as it was in the instant before interaction.
The direction of the primary neutrino is sampled uniformly from the allowed ranges in azimuth and cosine of zenith.
The event location is selected according to the injection mode, which is described in detail later in Section~\ref{sec:injection}.

\begin{table}[h!]
	\centering
    \begin{tabular}{r l  l}
        \toprule
        Parameter & Description & Allowed ranges \\
        \midrule
        $E_{\nu}^\texttt{min}$, $E_{\nu}^\texttt{max}$ & Neutrino injected energy & $ [\SI{100}\GeV,~\SI{1}\EeV]$ \\ 
        $\gamma$ & Spectral index power law & $  \left(-\infty,\infty\right)$ \\
        $\theta_{\nu}^\texttt{min}$, $\theta_{\nu}^\texttt{max}$ & Injected primary zenith angle & $\left[0,\pi\right]$ \\ 
        $\phi_{\nu}^\texttt{min}$, $\phi_{\nu}^\texttt{max}$ & Injected primary azimuth angle & $\left[0,2\pi\right]$ \\
        \bottomrule
    \end{tabular}
    \caption{\textbf{\textit{Flux Properties.}}
    Parameter names are given in the left column, their description in the center column, and values on the right column.
    These parameters are chosen with respect to the desired flux.
    The allowed energy range is driven by the extent of the provided cross section tables.
    }
    \label{tbl:injection_parameters}
\end{table}

The event kinematics are defined in terms of the Bjorken $x$ and Bjorken $y$ kinematic variables.
As shown in~\refeq{eq:bjorken}, Bjorken $y$ is the fractional energy carried away by the out-going lepton in a DIS interaction and Bjorken $x$ is the fraction of primary-particle momentum transferred by the weak interaction.
These two variables are given by
\begin{align}\label{eq:bjorken}
y&=1-\dfrac{E_{f}}{E_{i}} & x&=\dfrac{4 E_{i} E_{f}\sin^{2}\theta}{2m_{p}(E_{i}-E_{f})},
\end{align}
where $\theta$ is the angle between the trajectories of the initial- and final-state leptons, $m_{p}$ the proton mass, and $E_{i}$ and $E_{f}$ are the energies of the initial and final state lepton, respectively.

The Bjorken quantities are sampled, using b-splines from a joint 3D probability density function in the logarithm of each of $E_{i}$, $x$, and $y$ space, using the Metropolis-Hastings algorithm~\cite{MetHast:10.1093}.
Final state particle energies and deviations from the injected primary direction are then calculated analytically according to these kinematic variables.
Events are written to an \hdf~file as they are generated, and a \LeptonInjector{} Configuration (LIC) file is written in parallel storing the exact configuration settings used for generation including both differential and total cross sections used.
These LIC files are structured binary data files containing a header for meta-data, including a version number, and may evolve over time.
The backwards-compatibility of LIC files is of importance, and will be maintained in future versions of \LeptonInjector{} and \LeptonWeighter{}.
The structure of the \hdf~files and exact specifications of the LIC files are described in Appendices~\ref{sec:li_event} and~\ref{sec:lic_structure}, respectively. 

\subsection{Injection}\label{sec:injection}

The two modes for injecting events are ranged mode and volume mode; each accepts and requires the parameters described in \reftab{tbl:injection_defaults}.

In volume mode, a cylinder, oriented vertically, is constructed around the origin according to specified parameters and an interaction point is selected uniformly within that cylinder's volume.
This injection mode is suitable for simulating events which are approximately point-like for the purposes of detection, such as neutral-current interactions and charged-current $\nu_e$ interactions which produce particle showers which are fairly short in dense media compared to the size of the detector.

\begin{table}[]
    \begin{tabular}{l l c c }
        \toprule
        Parameter & Description & Defaults & IceCube Example\\
        \midrule
        \textit{InjectionRadius} & \makecell[l]{Max distance of closest approach\\from origin for injection}  & $\SI{1200}\m$ & $\SI{900}\m$ \\
        \textit{EndcapLength}    & \makecell[l]{Possible longitudinal extent of injection\\from point of closest approach} & $\SI{1200}\m$ & $\SI{900}\m$ \\[10pt]
        \textit{CylinderRadius}  & Radius of injection cylinder & $\SI{1200}\m$ & $\SI{700}\m$ \\
        \textit{CylinderHeight}  & Height of injection cylinder & $\SI{1200}\m$ & $\SI{1000}\m$ \\
        \bottomrule
    \end{tabular}
    \caption{\textbf{\textit{Detector Properties.}}
    Parameter names are given in the left column, their description in the middle-left column, and Defaults on the middle-right column. 
    These parameters are chosen to chosen in order to cover the detector. Example parameters for the IceCube Neutrino Observatory are provided on the far-right column.
    }
    \label{tbl:injection_defaults}
\end{table}

\begin{figure}[p]
    \centering
    \vspace{-3cm}
    \makebox[\linewidth]{
        \includegraphics[width=1.1\linewidth]{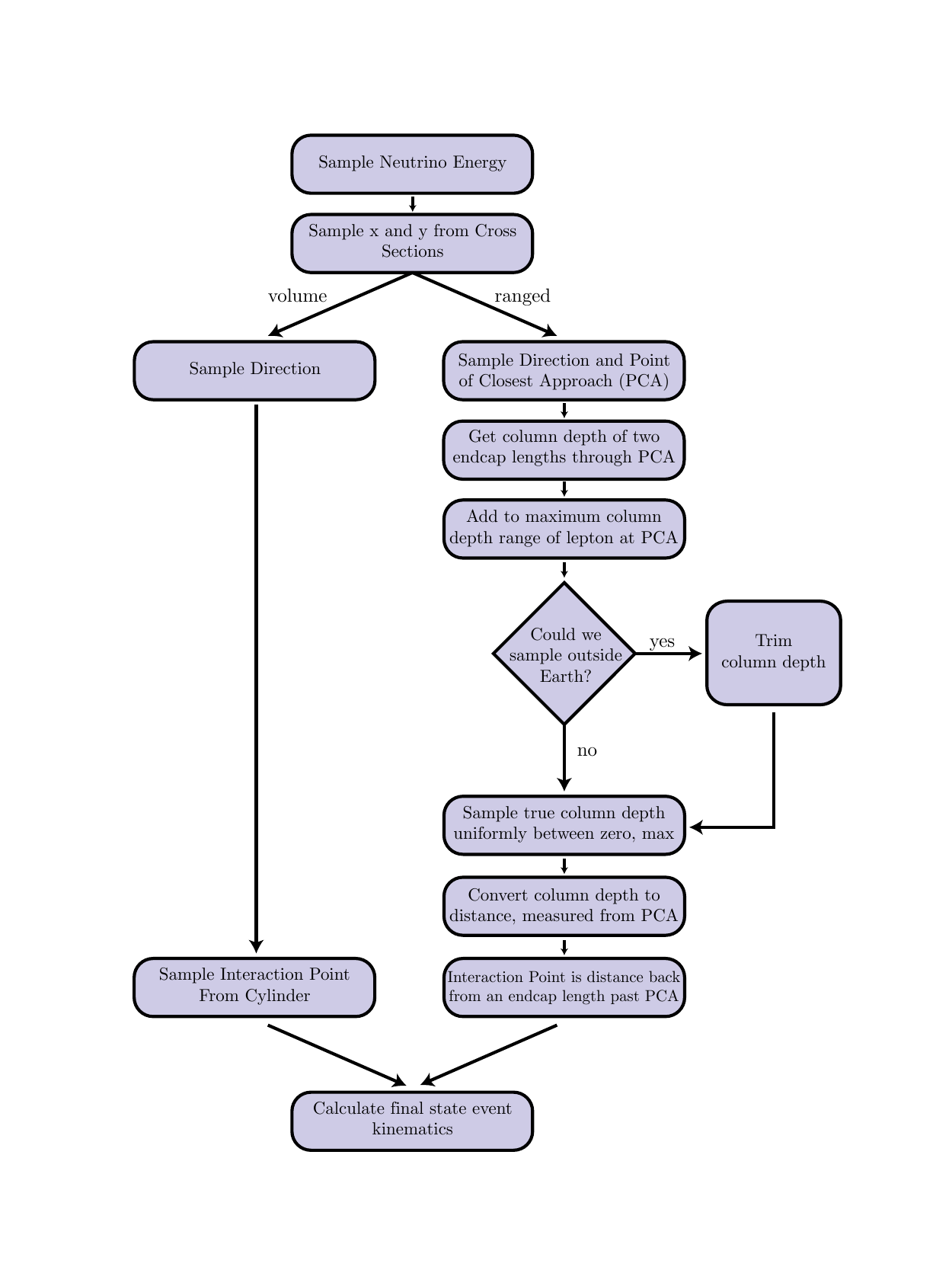}
    }
    \caption{A flowchart demonstrating the process of generating an event.}\label{fig:flowevent}
\end{figure}

\begin{figure}
    \centering
    \begin{subfigure}[b]{0.5\linewidth}
    	\centering
        \includegraphics[width=0.8\linewidth]{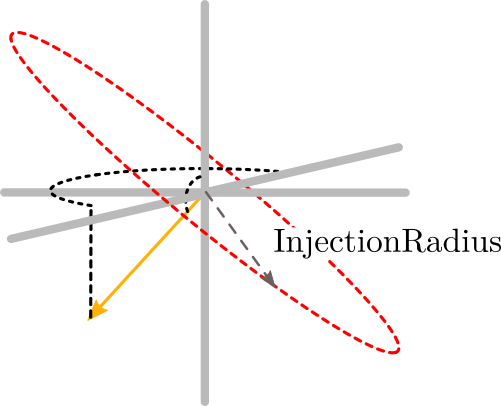}
        \caption{A direction (orange) is chosen and a perpendicular disc of radius \texttt{InjectionRadius} (red) is constructed.}
        \label{fig:lepton_ranged1}
    \end{subfigure}%
    \begin{subfigure}[b]{0.5\linewidth}
    	\centering
        \includegraphics[width=0.8\linewidth]{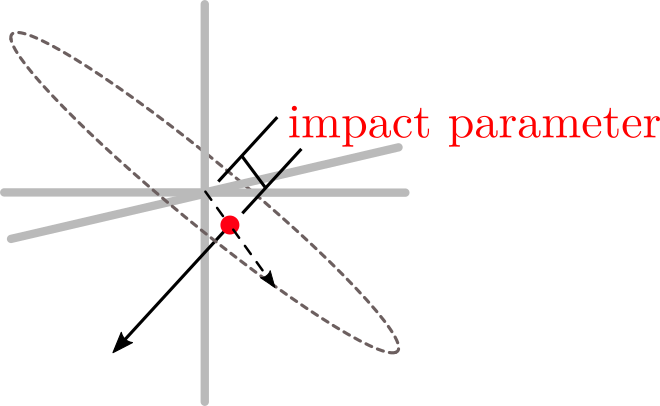}
        \caption{A point of closest approach is randomly sampled on that \texttt{InjectionRadius} disk. }
        \label{fig:lepton_ranged2}
    \end{subfigure} \\
    \begin{subfigure}[b]{0.5\linewidth}
    	\centering
        \includegraphics[width=0.8\linewidth]{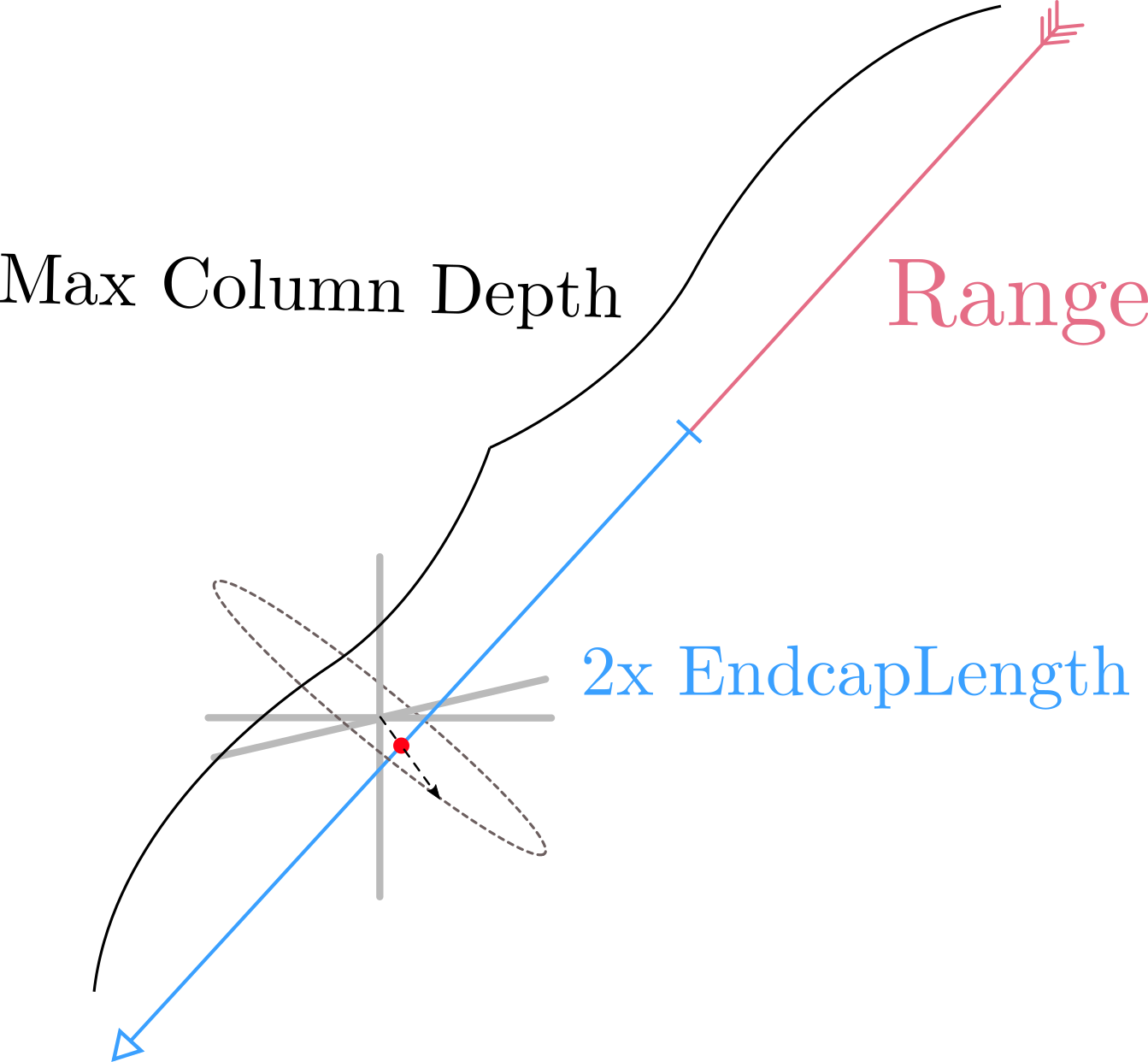}
        \caption{Preliminary column depth calculation of the lepton range (red) plus two \texttt{EndcapLength}s (blue)}
        \label{fig:lepton_ranged3}
    \end{subfigure}%
    \begin{subfigure}[b]{0.5\linewidth}
    	\centering
        \includegraphics[width=0.8\linewidth]{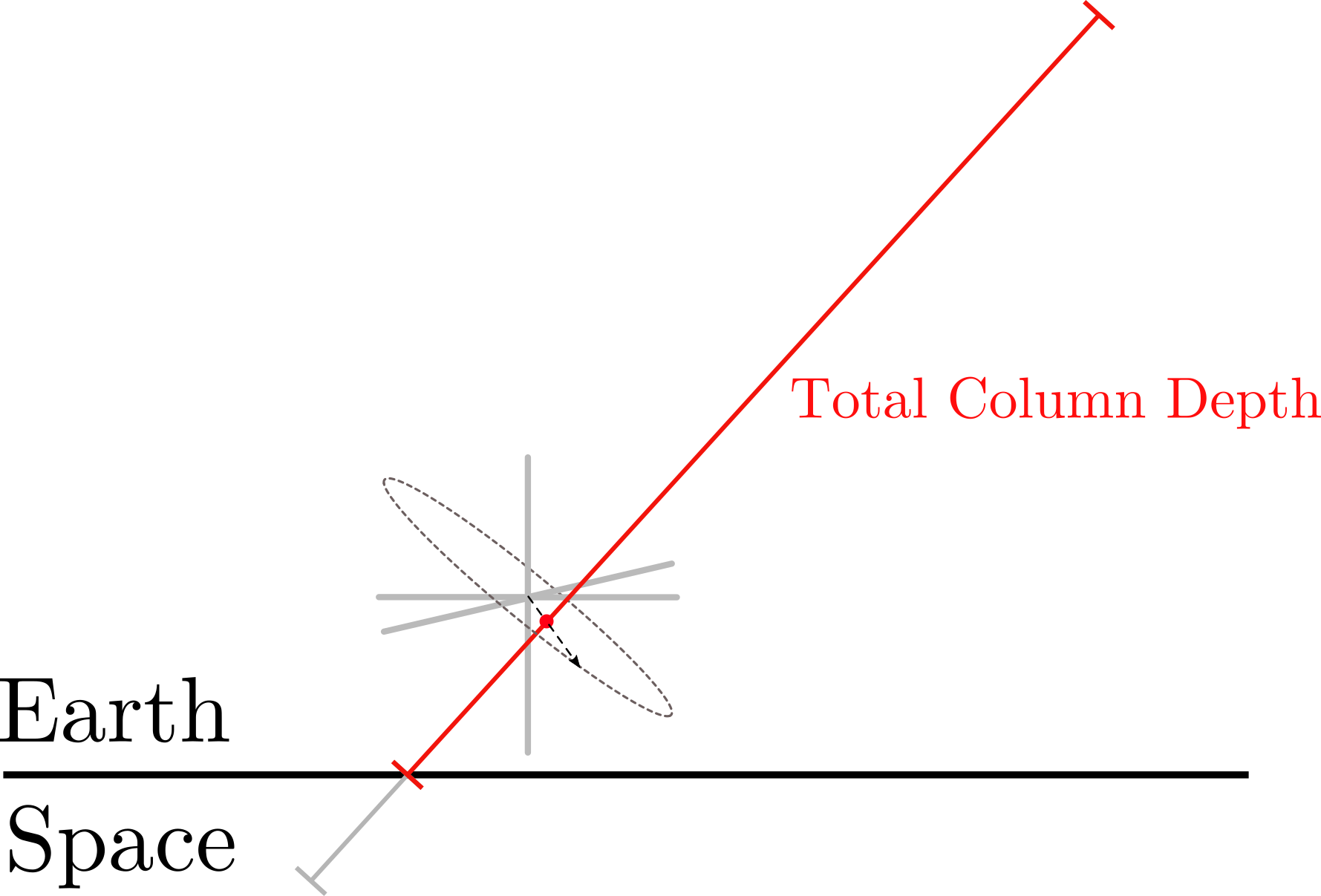}
        \caption{If the actual column depth available is less than the column depth computed in part (c), reduce it appropriately.}
        \label{fig:lepton_ranged4}
    \end{subfigure} \\
    \begin{subfigure}[b]{0.5\linewidth}
    	\centering
        \includegraphics[width=0.8\linewidth]{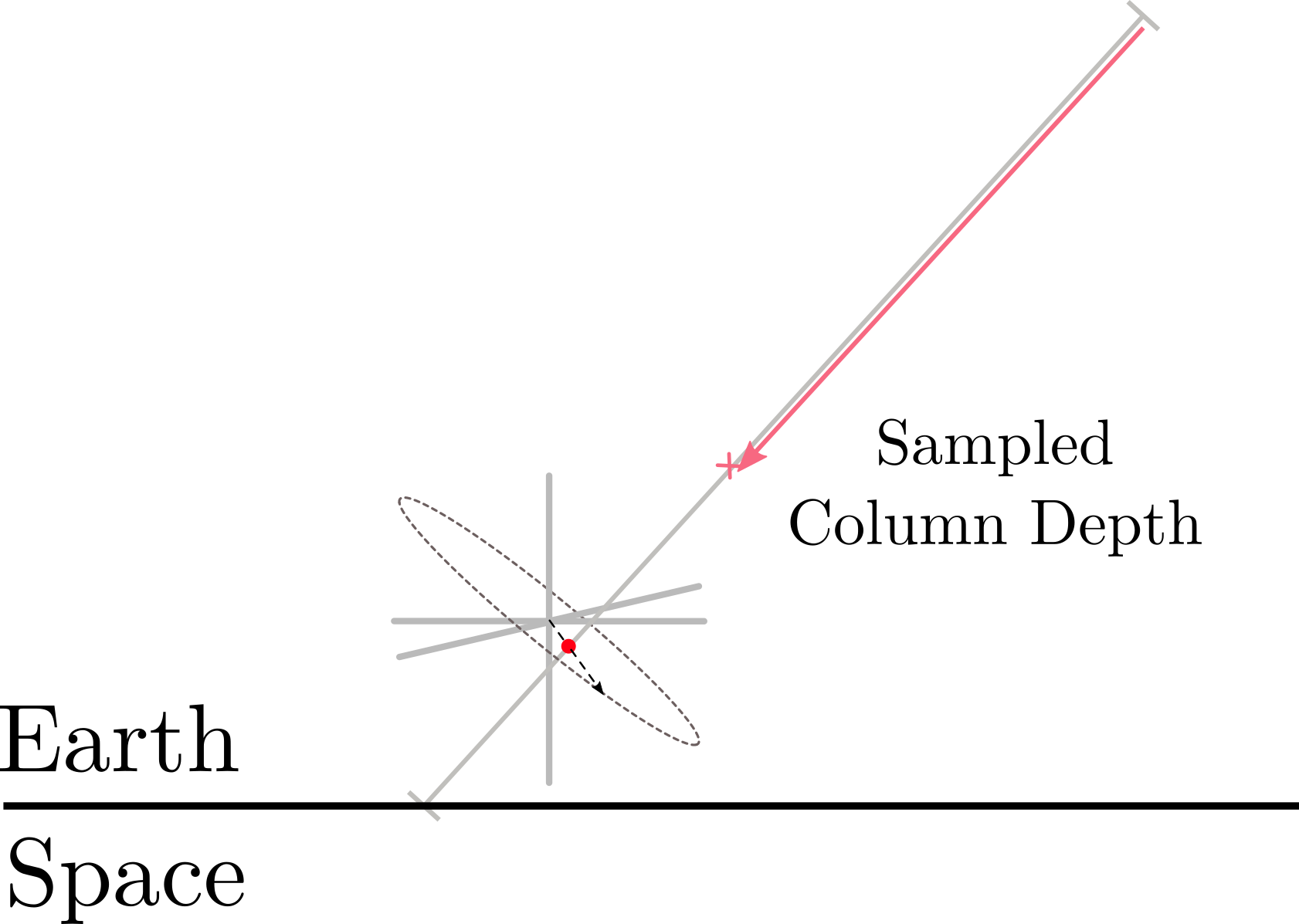}
        \caption{Sample uniformly from the total column depth available for insertion depth. }
        \label{fig:lepton_ranged5}
    \end{subfigure} \\
    \caption{Visualization of the ranged injection process and geometry.}
    \label{fig:lepton_ranged}
\end{figure}

The ranged mode process is illustrated in Figure~\ref{fig:lepton_ranged}.
This mode is intended as a counterpart of the volume injection mode.
Ranged mode is suitable for simulating events where the detection is due to visible daughter particles ($\mu^\pm$, $\tau^\pm$) which travel through dense media for distances comparable to or larger than the size of the detector.
It ensures sampling of interaction positions, over the whole volume of a target detector, both as far away as possibly visible to the detector due to daughter particles leaving the interaction, and proportional to local material density.
A typical example of an interaction type in this category is the charged-current $\nu_\mu$ interaction. 

During generation in ranged mode, a direction for the primary neutrino is first chosen within the allowed range of azimuth and zenith angles; this is shown in Figure~\ref{fig:lepton_ranged1}.
Then, as in Figure~\ref{fig:lepton_ranged2}, a point is randomly chosen from a disk of radius \textit{InjectionRadius} centered at the origin and perpendicular to the sampled direction; this point will be the point of closest approach (PCA) of the injected neutrino's projected path.
The distance from the sampled PCA to the origin is called the impact parameter.
Next, a range of possible positions along this path is determined in which the interaction position may be sampled.
This includes two `endcaps,' specified as lengths (\textit{EndcapLength}) on either side of the disk containing the PCA, and a maximum lepton `range.'
The endcaps are added to ensure that events are sampled over the entire volume of the detector, and the range is computed to account for the maximum distance that the charged lepton daughter of the interaction may travel.
This maximum distance is calculated according to 
\begin{align}
R_{\mu}(E) &= \dfrac{1}{d_{b}}\log\left(1 + E\dfrac{d_{a}}{d_{b}}\right), \text{ and} \label{eq:dima1} \\
R_{\tau}(E) &= R_{\mu}(E) + \left(3.8\times 10^{4}\right)\log\left(1+\tfrac{1}{5.6\times 10^{7}}\right) \label{eq:dima2},
\end{align}
with $d_{a}=0.212/1.2~[\si{\GeV\per\mwe}]$ and $d_{b}=(0.251\times 10^{-3})/1.2~[\si{\per\mwe}]$~\cite{chirkin2004propagating}.
$R_{\mu}$ and $R_{\tau}$ represent the maximum ranges, in meters water equivalent, for $\SI{99.9}\percent$ of muons and taus of energy $E$ in $\si\GeV$ respectively. 
The factor of 1.2 is to account for the observed deviations from the fits producing these max range functions. 
The maximum deviation was less than $\SI{20}\percent$, and as such $d_{b}$ is appropriately scaled.

The range of possible positions is converted to common units of column depth by taking into account the density of the material along the line formed by the two endcap lengths, including both local material around the detector and the Earth more generally, using a variation of the Preliminary Reference Earth Model~(PREM)~\cite{DZIEWONSKI1981297}, which we have extended with three uniform-density layers: a $\SI{2.6}\km$ thick clear-ice layer, a $\SI{200}\m$ thick firn layer, and a $\SI{103}\km$ atmosphere layer; these are demonstrated in Appendix~\ref{sec:earthdensity}.
These extra layers cover the entire Earth, and are required to accurately distribute the events, with respect to depth, in ranged mode.
This gives the preliminary maximum column depth within which the generator should ideally sample the interaction point.
The geometry of this calculation is shown in Figure~\ref{fig:lepton_ranged3}.
The model of the surrounding material is then integrated again, to determine whether the amount of column depth desired from the preliminary calculation actually exists along the path; at high energies it may not if the lepton range is sufficient to extend outside the outermost layer of the Earth model.
In this case, the maximum column depth is reduced to the physically available value, as in Figure~\ref{fig:lepton_ranged4}.
The resulting column depth is called the total column depth.
The amount of column depth the neutrino should traverse before interacting is then sampled uniformly between zero and this total column depth, and then converted to a physical position along the chosen path by a final integration of the material model as shown in Figure~\ref{fig:lepton_ranged5}.

\subsection{Comparisons\label{sec:compare}}

\begin{figure}
    \centering
    \begin{subfigure}[t]{0.5\linewidth}
    	\centering
        \includegraphics[width=\linewidth]{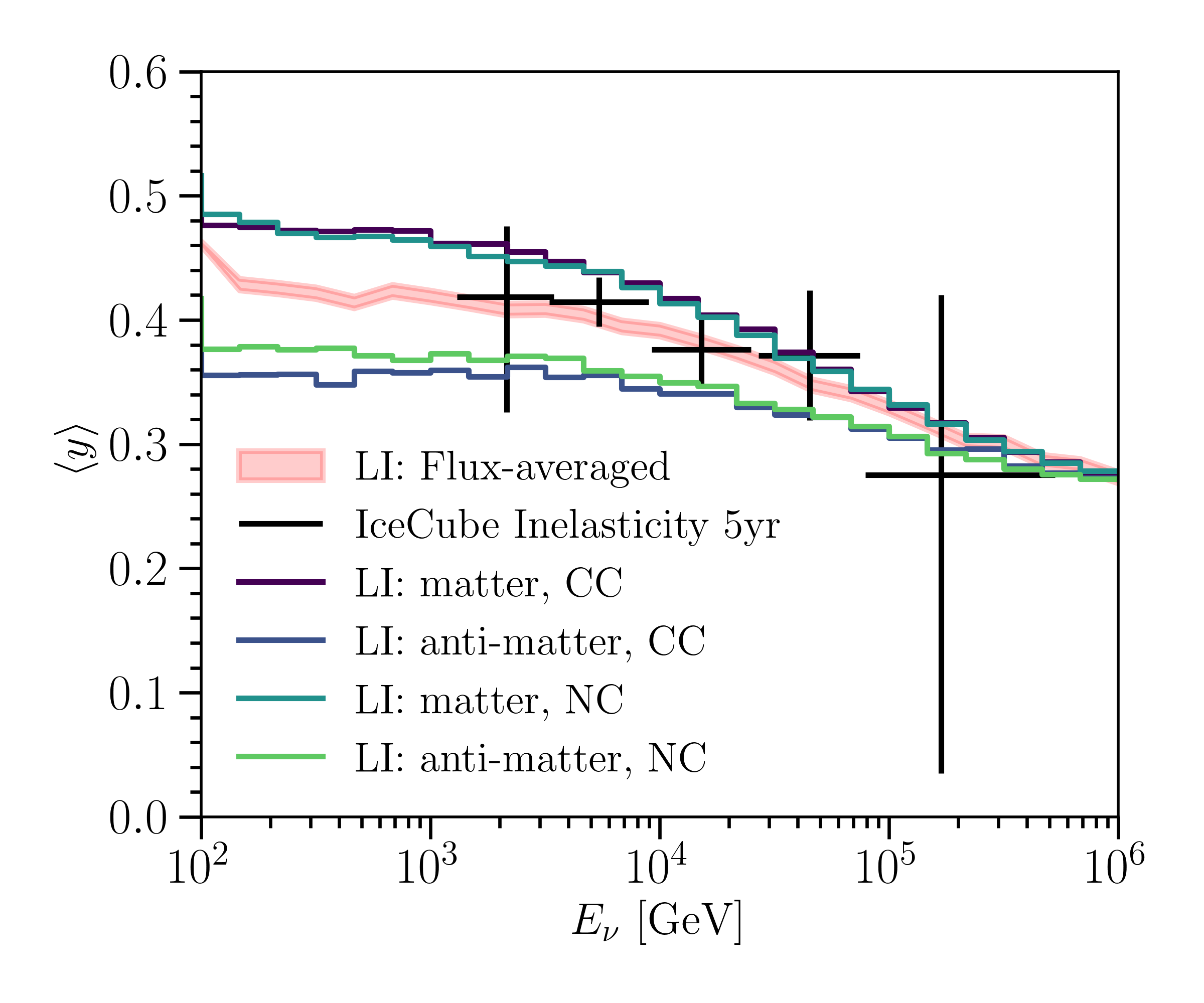}
        \caption{\LeptonInjector{} simulation with five years of IceCube inelasticity measurements overlaid.}\label{fig:bjorken_v_e}
    \end{subfigure}%
    \begin{subfigure}[t]{0.5\linewidth}
    	\centering
        \includegraphics[width=\linewidth]{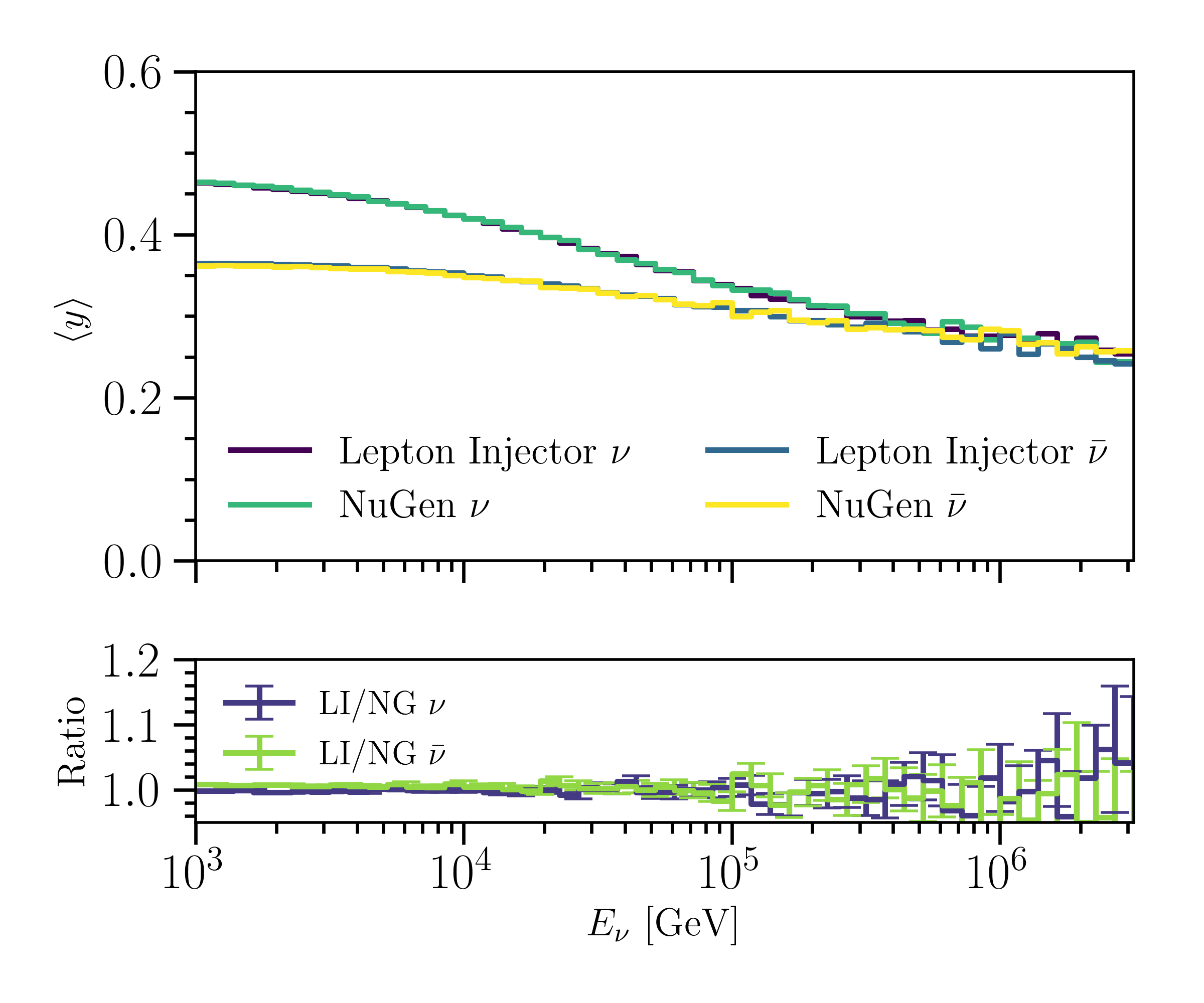}
        \caption{Top: $\langle y \rangle$ as a function of $E_{\nu}$, generated by \LeptonInjector and \NuGen. Note that trend lines are directly superimposed. Bottom: the ratio of $\langle y \rangle$ from \LeptonInjector{} and \NuGen.}\label{fig:nugen_li_bjorken}
    \end{subfigure}
    \caption{Average event inelasticity, with respect to neutrino energy in $\si\GeV$, for NC and CC events resulting from neutrinos and anti-neutrinos. }
\end{figure}

\begin{figure}
    \centering
    \begin{subfigure}[t]{0.5\linewidth}
    	\centering
        \includegraphics[width=0.85\linewidth]{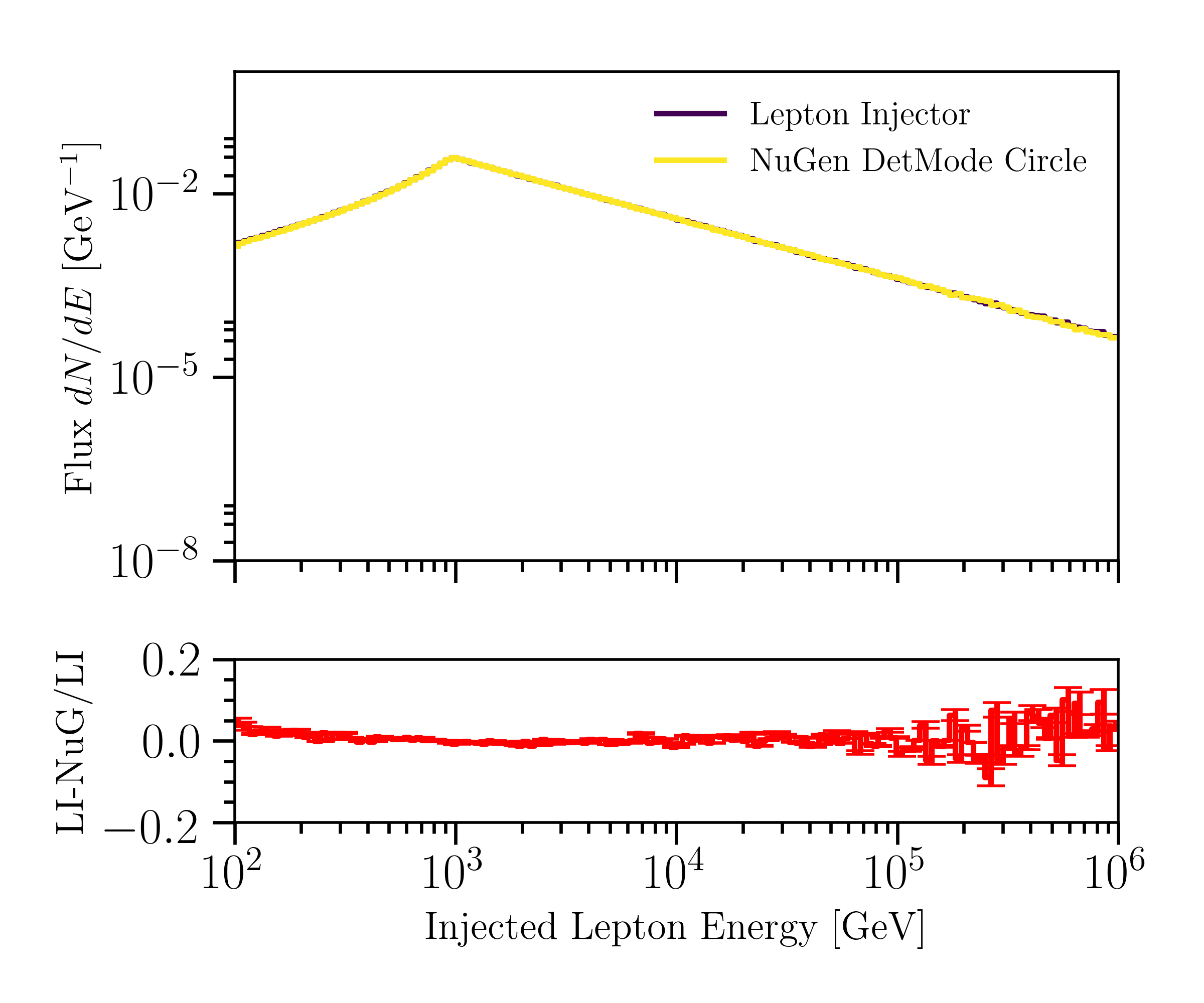}
        \caption{Top: injected lepton energy for an anti-neutrino events.\\Bottom: relative difference between \LeptonInjector{} and \NuGen{}.}
    \end{subfigure}%
    \begin{subfigure}[t]{0.5\linewidth}
    	\centering
        \includegraphics[width=0.85\linewidth]{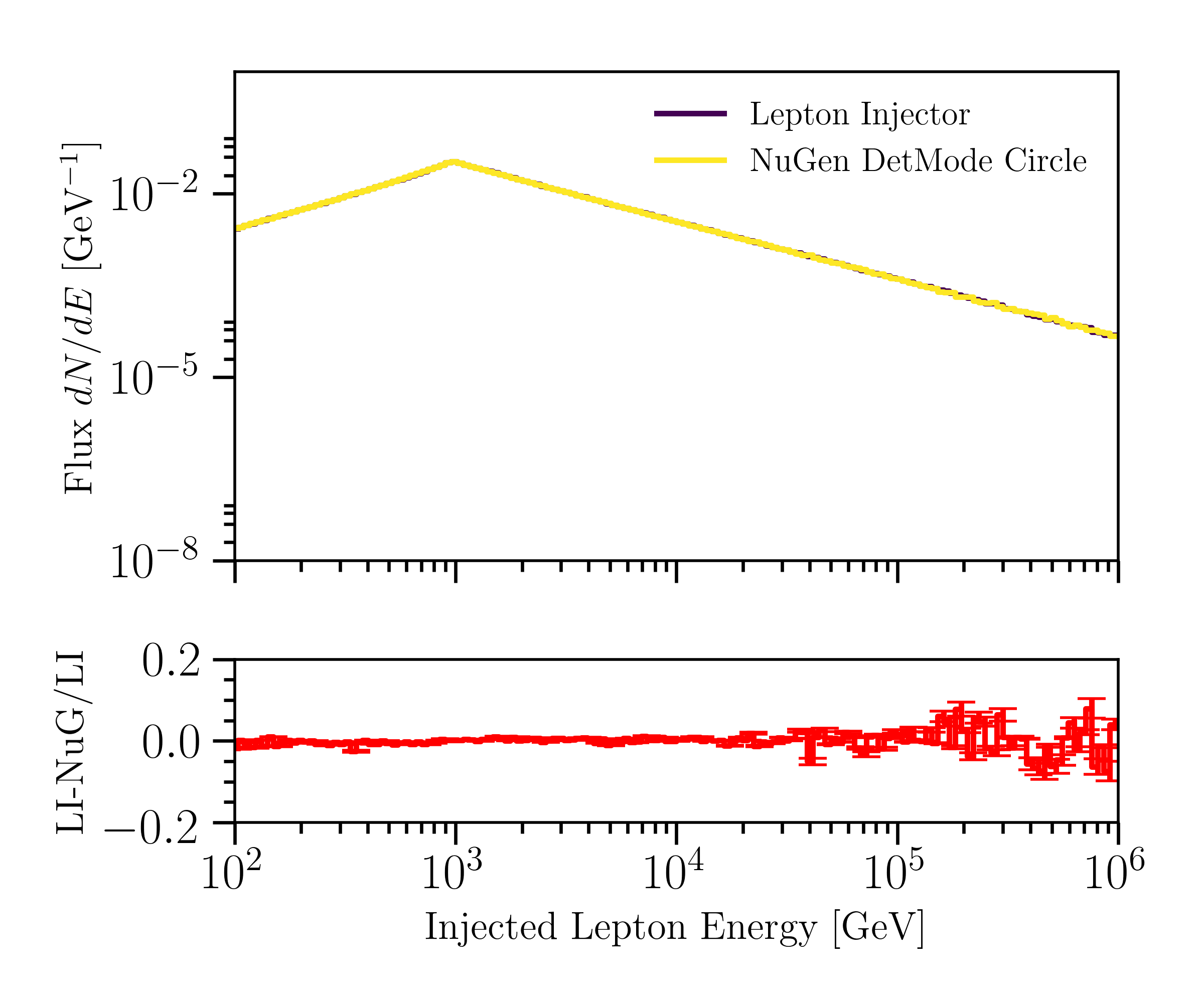}
        \caption{Top: injected lepton energy for neutrino events.\\Bottom: relative difference between \LeptonInjector{} and \NuGen{}.}
    \end{subfigure}
    \caption{Comparison between \LeptonInjector{} and \NuGen{}'s Detector Mode for the spectrum of the energies of leptons produced in the interactions. Total event energies were sampled with a $\SI{1}\TeV$ minimum. Note that the \LeptonInjector{} and \NuGen{} lines are directly superimposed.}\label{fig:secondary_energy}
\end{figure}

Prior to this generator, in IceCube, the primary neutrino event generator has been \NuGen{}. 
Similar to \LeptonInjector{}, \NuGen{} has been used to generate both up- and down-going neutrino events of all flavors and neutrino type; it has been used for the Monte Carlo event generation of numerous studies in IceCube and is thoroughly vetted.  
As part of the development of \LeptonInjector{}, comparisons were made between identical MC samples generated by \LeptonInjector{} and \NuGen{}'s `Detector Mode,' which uses an injection scheme roughly analogous to \LeptonInjector{}'s ranged mode. 
Events of all flavor, for both neutrino and anti-neutrino primaries, were generated for each neutrino type at a spectrum of $E^{-1}$ over all azimuth angles and up-going zenith angles.
A comparison of the spectra of injected lepton energies are shown in Figure~\ref{fig:secondary_energy}, and a comparison of the average inelasticity parameter as a function of primary neutrino energy is shown in Figure~\ref{fig:nugen_li_bjorken}. 
Additional comparisons were carried out for distributions of Bjorken $x$ and $y$, and the opening angle between injected particles, and were all found to be in agreement.

\LeptonInjector{} was also verified to produce events with a realistic distribution of final state kinematics.
Among other tests, to this end, we compared a large sample of generated events with the theoretical predictions of the final states resulting from CC, NC, and GR interactions, for which we use~\cite{CooperSarkar:2011pa} to model the DIS interactions and the analytical expressions given in~\cite{Glashow:1960zz,Gandhi:1995tf} for the GR.
Figure~\ref{fig:bjorken_v_e} compares the average inelasticity for an interaction of a given energy for neutral- and charged-current interactions with both neutrinos and anti-neutrinos; results from a five-year IceCube study on inelasticity distributions are overlaid~\cite{PhysRevD.99.032004} along with a flux-averaged \LeptonInjector{} sample.
This trend closely follows the predictions of~\cite{CooperSarkar:2011pa}; see~\cite{Binder:2017rlx} for an extended discussion.

\section{LeptonWeighter\label{sec:leptonweighter}}

The events produced by the \LeptonInjector{} algorithm described in Section~\ref{sec:overview} are generated at an arbitrary rate chosen by the user; \LeptonWeighter{} allows these events to then be re-weighted to any physical neutrino flux and interaction cross section.
Here, we briefly explain the reweighting procedure.

Suppose a sample of events was generated according to some ansatz distribution $\Phi(E)$, \textit{e.g.} according to
\begin{equation}
\frac{dN}{dE} = \Phi(E),
\end{equation}
where $E$ is the neutrino energy and $dN/dE$ is the expected flux density.
To re-weight the sample to a uniform distribution in energy, a weight $w_{\textrm{event}}$ is applied to each event, inversely proportional to the generating probability density:
\begin{equation}
w_{\textrm{event}}(E_{0}) = 1/\Phi(E_{0}),
\end{equation}
where $E_0$ is the energy of the event.
Suppose instead two sub-samples were generated from distributions $\Phi_a$ and $\Phi_b$, with the same domain in energy, and were then combined into one.
To re-weight events in the combined sample to a uniform distribution, a weight of 
\begin{equation}
w_{\textrm{event}}(E_{0}) =\dfrac{1}{ \Phi_a(E_{0}) + \Phi_b(E_{0}) }
\end{equation}
would be applied.
This is the generation weight, and accounts for the probability that either distribution could produce any given event. 
Events can then be re-weighted to a new distribution by evaluating their probability density in the new distribution and dividing by the generation weight. 

Extending this to \LeptonInjector{}, the probability density that a given \textbf{Generator} could have produced an event for each event is
\begin{equation}
p_\mathrm{MC} = N_\textrm{gen}\frac{1}{\Omega_\mathrm{gen} A_\mathrm{gen}} \times
\frac{\rho_\textrm{gen}(\ell)}{X_\textrm{gen}^\textrm{col}} \times \frac{1}{\sigma_\mathrm{tot}}\frac{\partial^2\sigma}{\partial x\partial y}\times \frac{\Phi(E)}{\int_{E_\mathrm{min}}^{E_\mathrm{max}}\Phi(E) dE}
\end{equation}
where $\Omega_{\mathrm{gen}}$ is the solid angle over which events were generated, $\Phi(E)$ is the power-law flux spectrum of the \textbf{Generator}, $A_{\mathrm{gen}}$ is the integrated area of the sampling surface, $\rho_\textrm{gen}(\ell)$ is the local mass density of targets, $X_\textrm{gen}^\mathrm{col}$ is total column depth of targets in the sampling region, $N_{\textrm{gen}}$ is the total number of generated events, and $\partial_{xy}\sigma$ and $\sigma_{\mathrm{total}}$ are the differential and total cross sections evaluated for the event, respectively.
As a result, $p_{\mathrm{MC}}$ has units of $\si{\per\steradian\per\cubic\cm\per\GeV}$.
For a single MC generator, whose exact definitions are discussed in~\cite{Gainer:2014}, the generation weight is the inverse of the generation probability density:
\begin{equation}
w_{\rm gen} = \frac{1}{p_{\mathrm{MC}}}.
\end{equation}
In the regime of small neutrino interaction probability, an event's final weight, in units of $\si{\per\s}$, is approximately given by
\begin{equation}
w_{\rm event} = \sum\limits_{\lbrace \textrm{gen}\rbrace} \underbrace{\left( \frac{X_\textrm{physical}^\mathrm{col} N_{A}}{M_\textrm{target}} \times \frac{\rho_\textrm{physical}(\ell)}{X_\textrm{physical}^\textrm{col}} \times \frac{\partial^2\sigma}{\partial x \partial y} \times\Phi_{\mathrm{physical}} \right) }_{\text{physical distribution}} \times \underbrace{ w_{\mathrm{gen}}}_{\text{gen weight}},
\label{eq:approx_weight}
\end{equation}
where $\{\textrm{gen}\}$ indicates the set of generators, $M_\textrm{target}$ is the molar mass of the target, $\Phi_{\mathrm{physical}}$ is the desired physical flux of neutrinos at the detector, $N_A$ is Avogadro's constant, $X_\textrm{physical}^\textrm{col}$ is calculated as
\begin{equation}
X_\textrm{physical}^\textrm{col} = \int\limits_{\ell_{i}}^{\ell_{f}}\rho_\textrm{physical}(\ell) d\ell
\end{equation}
along the path $\ell$ the particle would take to interact, and $\partial_{xy}\sigma$ is the differential cross section evaluated for the event.
Note that $X_\textrm{physical}^\textrm{col}$ is the physical column density between the generation boundaries $\ell_f$ and $\ell_i$, which are not necessarily the same across different generators.
If the physical and generation density models are the same, then the ${\rho_\textrm{physical}(\ell)}/{X_\textrm{physical}^\textrm{col}}$ and ${\rho_\textrm{gen}(\ell)}/{X_\textrm{gen}^\textrm{col}}$ terms cancel.
For a more complete description of the weighting that covers non-negligible interaction probabilities, see Appendix~\ref{sec:weight_append}.

This weighting calculation procedure is implemented in the \LeptonWeighter{} {\ttf C++} library and Python module; it is available in~\cite{LeptonWeighterRepository}.
\LeptonWeighter{} uses the LIC files generated by \LeptonInjector{} to calculate the above generation weights, then a user-specified cross section and flux to calculate event weights.

\begin{figure}[h]
 \centering
    \begin{subfigure}[b]{0.45\linewidth}
    	\centering
        \includegraphics[width=1.0\linewidth]{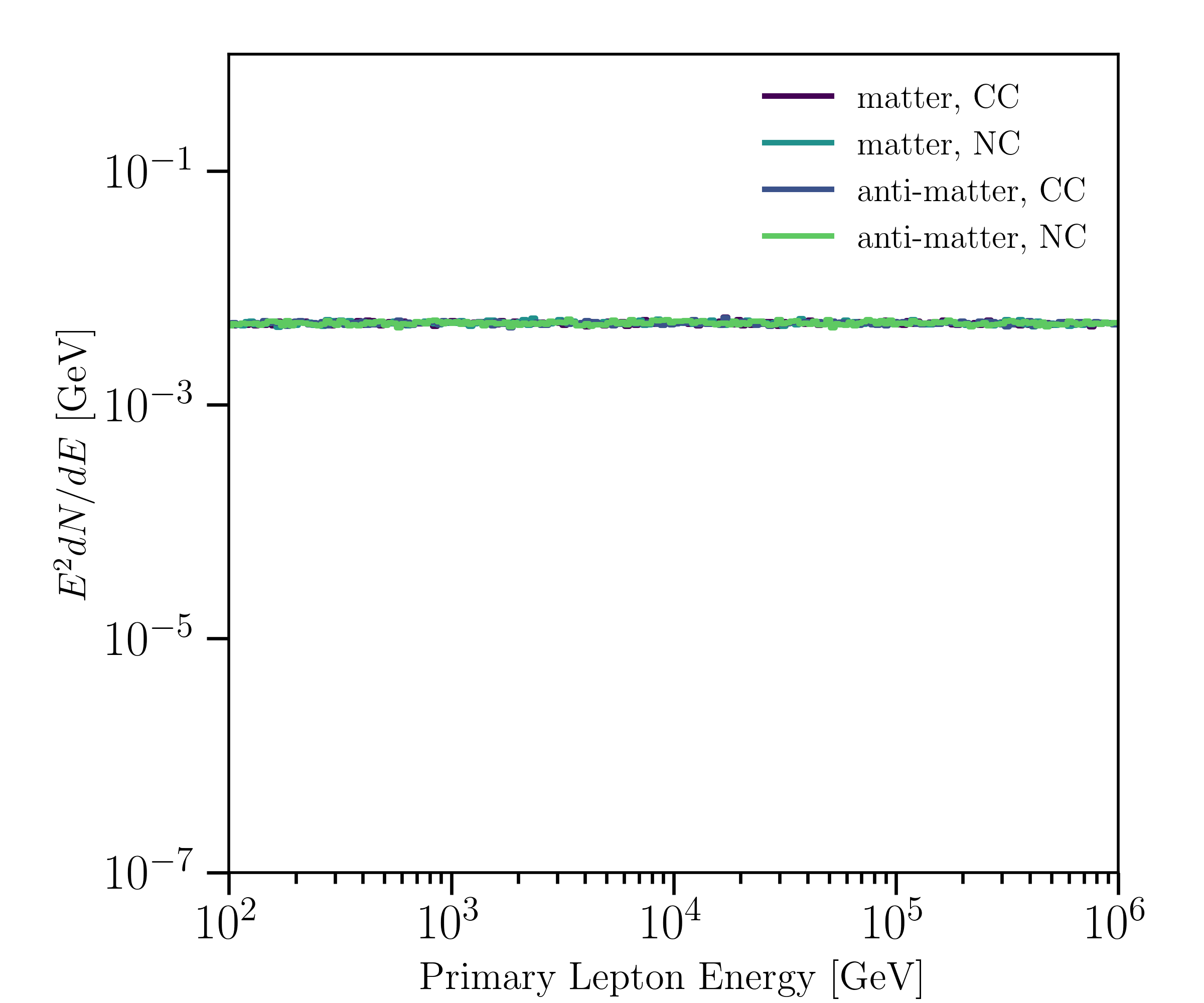}
        \caption{Unweighted}
    \end{subfigure}%
    \begin{subfigure}[b]{0.45\linewidth}
    	\centering
        \includegraphics[width=1.0\linewidth]{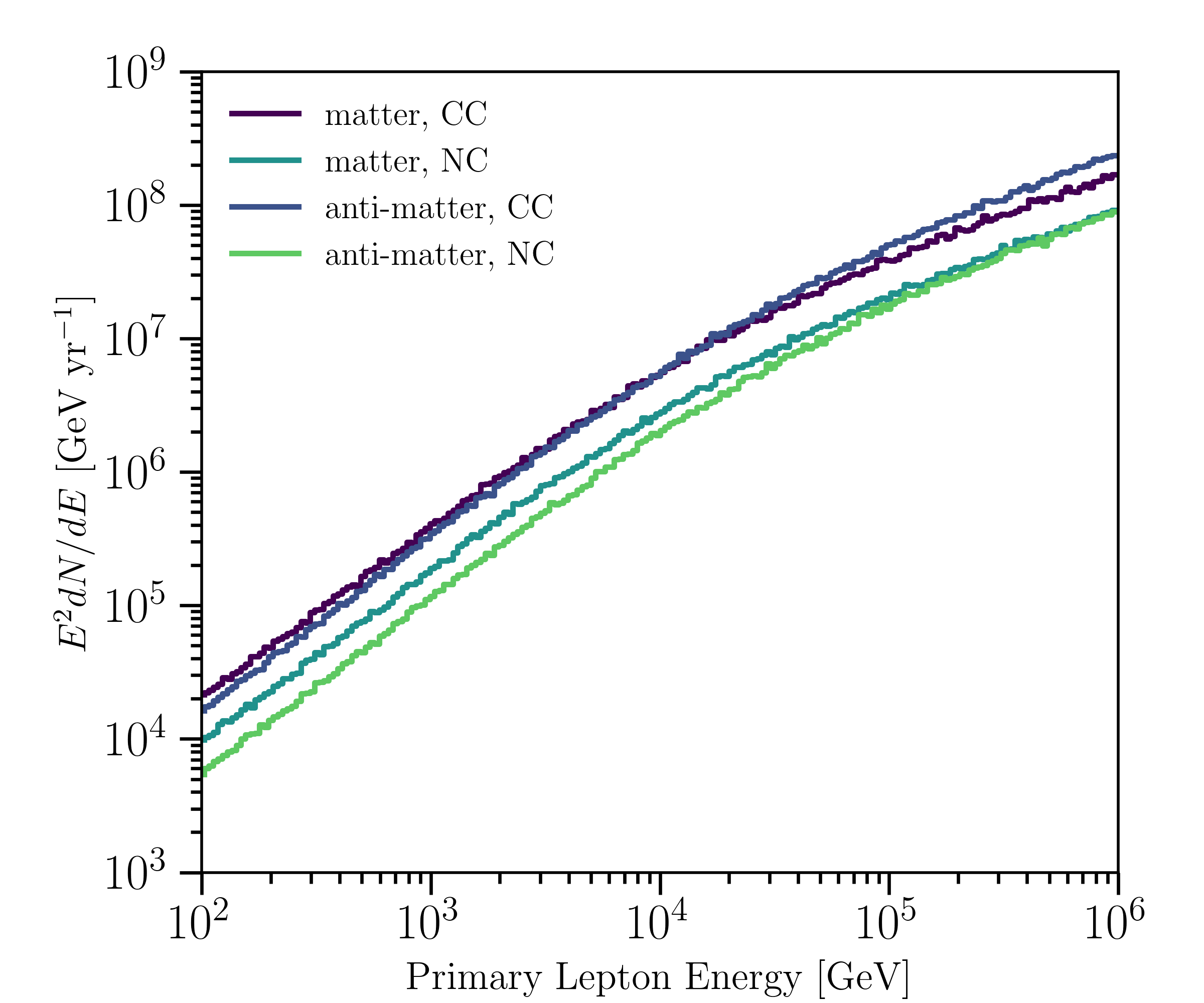}
        \caption{$E^{-2}$ astrophysical flux convolved with DIS cross section}
    \end{subfigure}\\
    \begin{subfigure}[b]{0.45\linewidth}
    	\centering
        \includegraphics[width=1.0\linewidth]{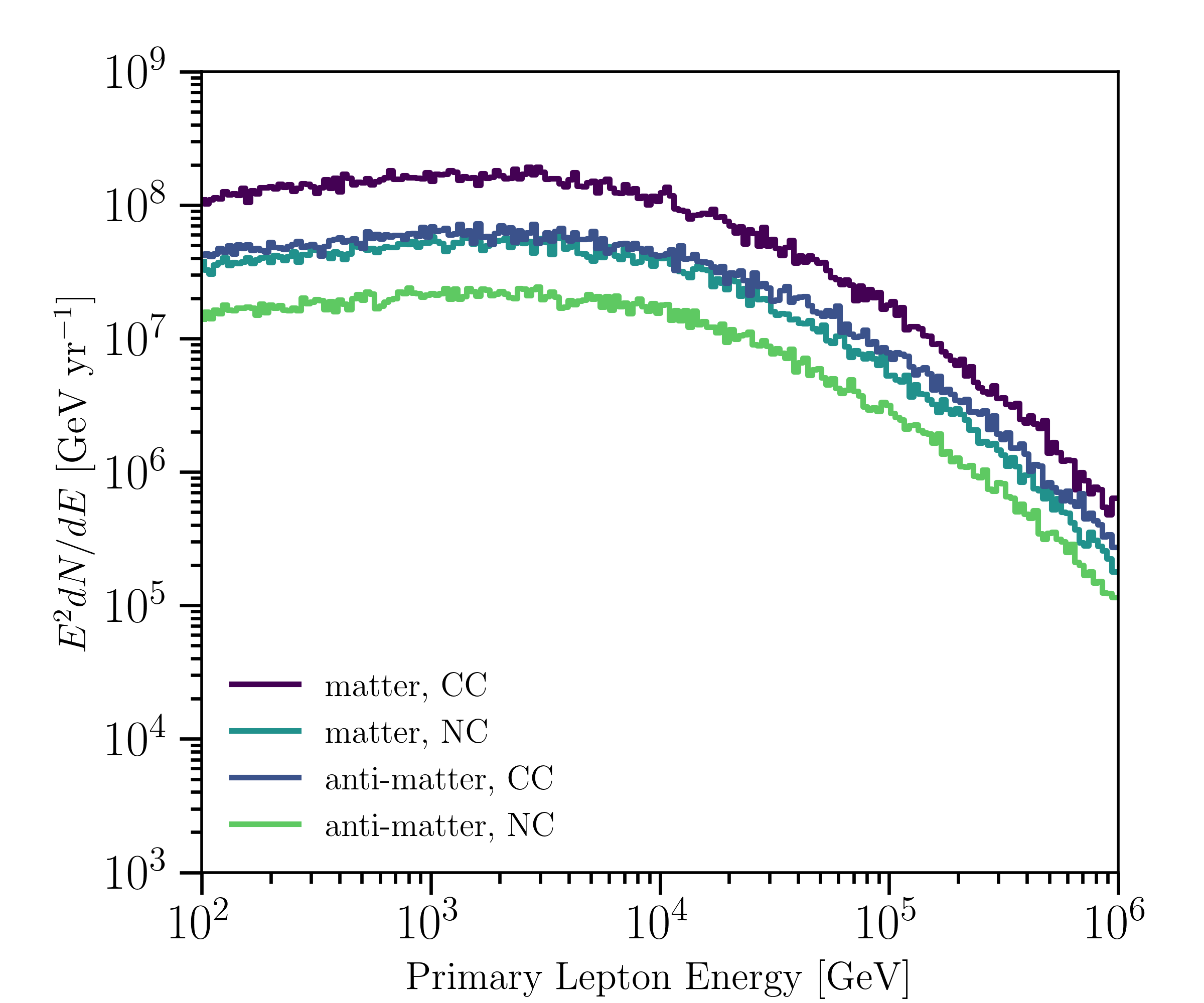}
        \caption{Atmospheric flux convolved with DIS cross section}
    \end{subfigure}%
    \begin{subfigure}[b]{0.45\linewidth}
    	\centering
        \includegraphics[width=1.0\linewidth]{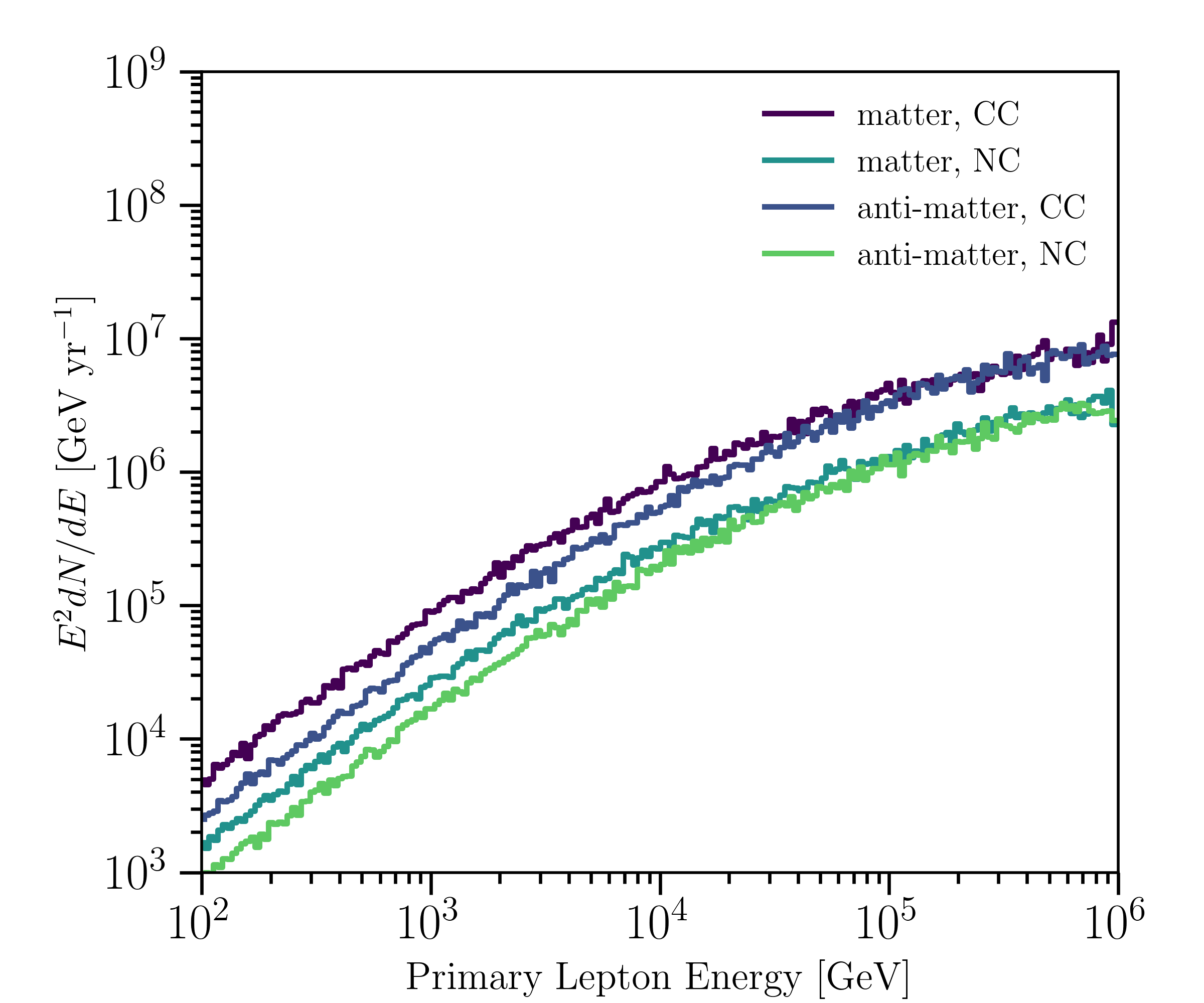}
        \caption{atmospheric flux with additional NSI}
    \end{subfigure}
    \caption{Several re-weightings of the same sample with one year of live-time. Top-left: unweighted, top-right: re-weighted to an $E^{-2}$ astrophysical flux, bottom-left: re-weighted to an atmospheric flux, bottom-right: re-weighted to a sample with non-standard interactions used propagation of atmospheric neutrinos. Here an extra, lepton-number violating, non-diagonal, term has been added to the neutrino propagation Hamiltonian.}\label{fig:weightings}
\end{figure}

\subsection{Code Structure\label{sec:lw_code}}

\begin{figure}[p]
    \centering
    \makebox[\linewidth]{
        \includegraphics[width=1.2\linewidth]{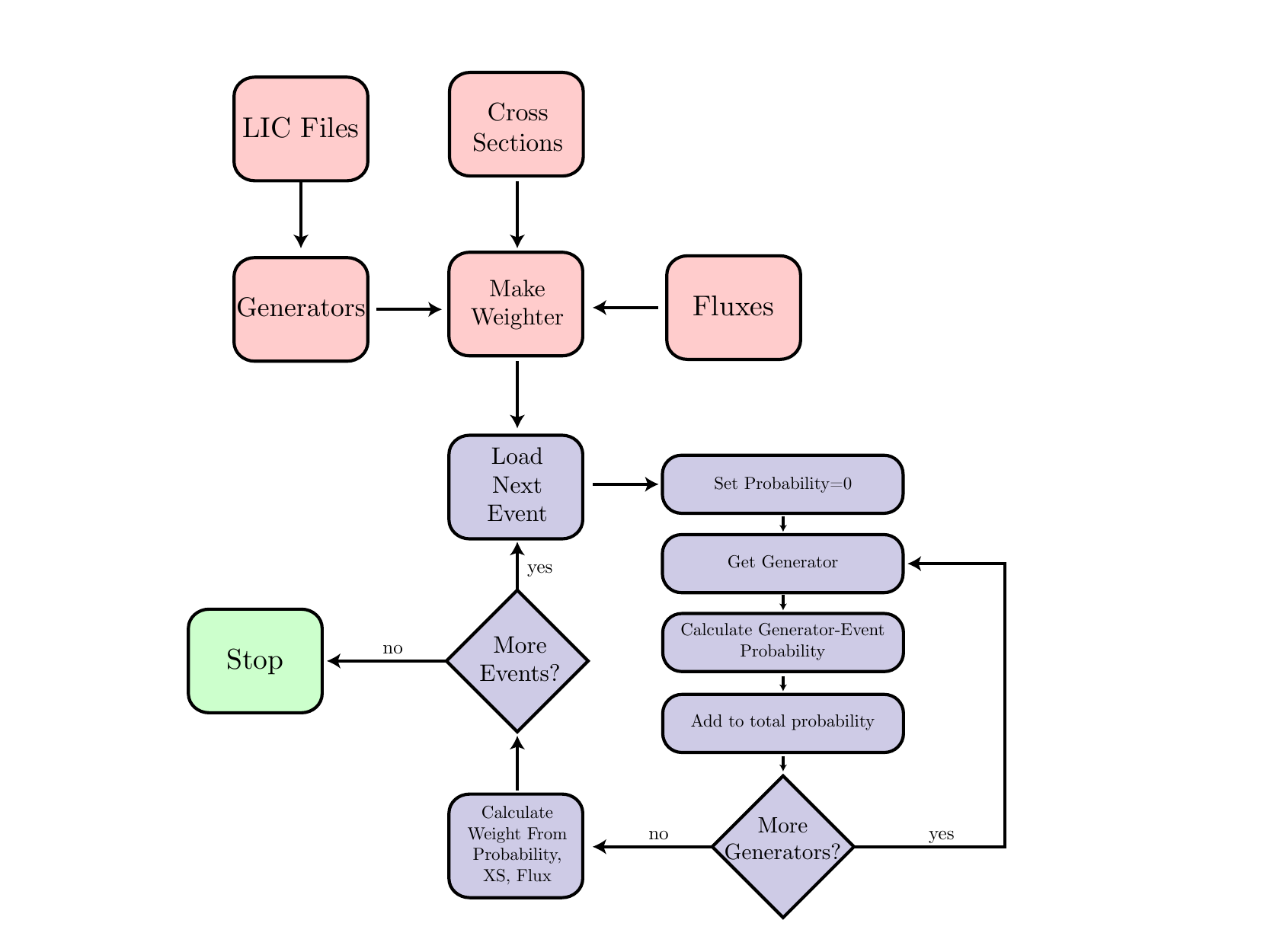}
    }
    \caption{A flowchart illustrating the process for calculating the individual weights of a collection of events.}
    \label{fig:LWflow}
\end{figure}

\LeptonWeighter{} divides its functionality into distinct components: fluxes, cross sections, \textbf{Generators}, and \textbf{Weighters}. 
\textbf{Flux} objects are constructed to define a flux to which a sample should be weighted.
\textbf{CrossSection} objects are similarly constructed with paths to locally saved FITS files of the same format as those used by \LeptonInjector{}, and are used to define the cross sections used to weight the sample's events.
\LeptonWeighter{} constructs \textbf{Generators} by reading LIC files from disk and deserializing them.
These \textbf{Generators} contain the exact simulation parameters used by \LeptonInjector{} to generate events, and are able to calculate the generation weight for any event following a process illustrated in Figure~\ref{fig:LWflow}.

\LeptonWeighter{} creates a \textbf{Weighter} object by using a list of \textbf{Generators}, a \textbf{Flux} object, and a \textbf{CrossSection} object.
The \textbf{Weighter} object uses an event's properties, see Figure~\ref{fig:tiks_struct} in the Appendix, to calculate a weight as defined by Eq.~\eqref{eq:approx_weight}. 
Figure~\ref{fig:weightings} demonstrates an all-flavor Monte Carlo sample composed of equal parts neutral- and charged-current DIS events, generated using an $E^{-2}$ spectrum, and re-weighted to multiple different fluxes.
The top-left plot shows the unweighted sample; the top-right left plot shows the sample reweighed to an astrophysical flux and weighted to the CSMS calculation of the DIS cross section~\cite{CooperSarkar:2011pa}; the bottom-left is re-weighted to an atmospheric neutrino flux and convolved with the same DIS cross sections~\cite{CooperSarkar:2011pa}; the bottom-right plot shows the sample re-weighted to an atmospheric neutrino flux, with non-standard neutrino interaction (NSI) parameter strength of $\varepsilon_{\mu\tau}=2 \times 10^{-1}$.
See~\cite{Dev:2019anc} for a precise definition of this parameter and nuSQuIDS~\cite{Arguelles:2020hss,arguelles:2015nu} for the NSI implementation used.

In general, the weighting scheme allows to modify an already generated Monte Carlo set to any cross section that maps onto the same final states. 
For example, the DIS cross section could be a perturbative QCD calculation such as the CSMS@NLO~\cite{CooperSarkar:2011pa} or the BGR@NNLO calculation given in~\cite{Bertone:2018dse} or a phenomenological estimate using the colour-dipole model~\cite{arguelles:2015nu}. 
Similarly for the Glashow process one could use the original tree-level calculation in~\cite{Glashow:1960zz} or the updated calculation including radiative corrections given in~\cite{Gauld:2019pgt}.

\section{Broader applications}
Although \LeptonInjector{} and \LeptonWeighter{} have principally been developed for use by IceCube, the injection and weighting techniques are broadly applicable for experiments that need to simulate natural sources of neutrinos above $\SI{10}\GeV$.
To adapt the software to other experiments we must account for differences in detector geometry and material composition within and around the detector.

The size of the injection region can be easily adjusted by changing the \textit{InjectionRadius}, \textit{EndcapLength}, \textit{CylinderRadius}, and \textit{CylinderHeight} parameters to encompass the extent of the detector.
As long as the detector occupies a significant fraction of this cylindrical volume, the event injection remains efficient.

The material model used by \LeptonInjector{} and \LeptonWeighter{} has a simple implementation that models the Earth as a series of cylindrical shells with a radially varying polynomial density distribution.
The polar ice cap is modeled as an offset spherical shell, also with a radially varying polynomial density.
This implementation works well for detectors embedded in media that conform to spherical symmetry such as KM3NeT~\cite{Adrian-Martinez:2016fdl} in the Mediterranean Sea.
However, this approach to the material model breaks down when the symmetry is broken, as is the case for GVD in Lake Baikal~\cite{Avrorin:2018ijk} and the $\SI{17}\kt$ liquid Argon modules planned for DUNE~\cite{Abi:2020loh}.
To accommodate these experiments a more detailed software model of the surrounding material would need to be implemented.
As long as the injection and weighting procedures query the updated material mode, no other modifications should be necessary.

\section{Conclusions\label{sec:conclusions}}

Here we have presented the first publicly available neutrino telescope event generator for $\si\GeV$-$\si\PeV+$ energy ranges that factorizes the problem of Earth and atmospheric propagation from the event generation.
The valid energy range of the generator is not limited by the software, but by the input cross sections provided by the user. 
The default CSMS cross section~\cite{CooperSarkar:2011pa} provided with the code has less than $\SI{5}\percent$ uncertainty in the $\SI{100}\GeV$ to $\SI{100}\EeV$ energy range.
The factorization allows for streamlined and efficient production of neutrino events. 
\LeptonInjector{}, along with its sister software \LeptonWeighter{}, satisfy the needs of generating events for gigaton-scale neutrino observatories.
The current implementation contains the most significant processes relevant to current analyses performed by these observatories, however we expect that this work will be extended as new calculations are made available and newer experimental needs arise.
To this end, the code discussed in this paper follows an open-source model. 
Improvements recently proposed in the literature include: adding sub-leading neutrino interactions such as interactions with the nuclear coulomb field~\cite{Seckel:1997kk,Alikhanov:2014uja,Zhou:2019vxt,Beacom:2019pzs}, which is expected to be a $\SI{10}\percent$ contribution at $\SI{1}\PeV$; use of updated DIS models such as those given in~\cite{Garcia:2020jwr}; inclusion of trident neutrino events~\cite{Altmannshofer:2014pba,Ge:2017poy,Ballett:2018uuc,Altmannshofer:2019zhy,Beacom:2019pzs}; inclusion of nuclear effects on interactions~\cite{Bertone:2018dse}; inclusion of new physics processes such as production of heavy-neutral leptons~\cite{PhysRevLett.119.201804,Magill:2018jla,Cline:2020mdt} or dark neutrinos~\cite{Bertuzzo:2018itn,Blennow:2019fhy,Ballett:2019cqp,Coloma:2019qqj,Abdullahi:2020nyr}; inclusion of new neutrino interactions mediated by light $Z$-prime~\cite{Cherry:2016jol,Bakhti:2018avv,Ballett:2019xoj}; among others.

\section*{Acknowledgements}
The IceCube collaboration acknowledges the significant contributions to this manuscript from Carlos Arg\"uelles, Austin Schneider, and Benjamin Smithers.
We acknowledge the support from the following agencies:
USA {\textendash} U.S. National Science Foundation-Office of Polar Programs,
U.S. National Science Foundation-Physics Division,
Wisconsin Alumni Research Foundation,
Center for High Throughput Computing (CHTC) at the University of Wisconsin{\textendash}Madison,
Open Science Grid (OSG),
Extreme Science and Engineering Discovery Environment (XSEDE),
Frontera computing project at the Texas Advanced Computing Center,
U.S. Department of Energy-National Energy Research Scientific Computing Center,
Particle astrophysics research computing center at the University of Maryland,
Institute for Cyber-Enabled Research at Michigan State University,
and Astroparticle physics computational facility at Marquette University;
Belgium {\textendash} Funds for Scientific Research (FRS-FNRS and FWO),
FWO Odysseus and Big Science programmes,
and Belgian Federal Science Policy Office (Belspo);
Germany {\textendash} Bundesministerium f{\"u}r Bildung und Forschung (BMBF),
Deutsche Forschungsgemeinschaft (DFG),
Helmholtz Alliance for Astroparticle Physics (HAP),
Initiative and Networking Fund of the Helmholtz Association,
Deutsches Elektronen Synchrotron (DESY),
and High Performance Computing cluster of the RWTH Aachen;
Sweden {\textendash} Swedish Research Council,
Swedish Polar Research Secretariat,
Swedish National Infrastructure for Computing (SNIC),
and Knut and Alice Wallenberg Foundation;
Australia {\textendash} Australian Research Council;
Canada {\textendash} Natural Sciences and Engineering Research Council of Canada,
Calcul Qu{\'e}bec, Compute Ontario, Canada Foundation for Innovation, WestGrid, and Compute Canada;
Denmark {\textendash} Villum Fonden, Danish National Research Foundation (DNRF), Carlsberg Foundation;
New Zealand {\textendash} Marsden Fund;
Japan {\textendash} Japan Society for Promotion of Science (JSPS)
and Institute for Global Prominent Research (IGPR) of Chiba University;
Korea {\textendash} National Research Foundation of Korea (NRF);
Switzerland {\textendash} Swiss National Science Foundation (SNSF);
United Kingdom {\textendash} Department of Physics, University of Oxford.
United Kingdom {\textendash} Science and Technology Facilities Council (STFC), part of UK Research and Innovation.

\bibliographystyle{elsarticle-num}
\bibliography{main.bib}
\clearpage

\appendix

\ifx \standalonesupplemental\undefined
\setcounter{page}{1}
\setcounter{figure}{0}
\setcounter{table}{0}
\setcounter{equation}{0}
\fi

\renewcommand{\thepage}{Supplemental Material-- S\arabic{page}}
\renewcommand{\figurename}{SUPPL. FIGURE}
\renewcommand{\tablename}{SUPPL. TABLE}

\renewcommand{\theequation}{A\arabic{equation}}
\clearpage

\section{\LeptonInjector{} Event Structure~\label{sec:li_event}}

\begin{figure*}[htb!]
\centering
\begin{tikzpicture}[
  grow via three points={one child at (0.5,-0.7) and
  two children at (0.5,-0.7) and (0.5,-1.4)},
  edge from parent path={(\tikzparentnode.south) |- (\tikzchildnode.west)}]
  \node {MC Events}
    child { node [selected] {\ttf properties}
        child { node [label=right:{energy of the interacting neutrino. [$\si\GeV$]}] {\ttf totalEnergy}}	
        child { node [label=right:{initial zenith of the lepton. [$\si\radian$]}] {\ttf Zenith}}
        child { node [label=right:{initial azimuth of the lepton. [$\si\radian$]}] {\ttf Azimuth}}
        child { node [label=right:{Bjorken $x$ of the interaction. [dimensionless]}] {\ttf finalStateX }}	
        child { node [label=right:{Bjorken $y$ of the interaction. [dimensionless]}] {\ttf finalStateY}}
        child { node [label=right:{Code of first final state particle. [dimensionless]}] {\ttf finalType1}}
        child { node [label=right:{Code of second final state particle. [dimensionless]}] {\ttf finalType2}}
        child { node [label=right:{Code of parent particle. [dimensionless]}] {\ttf initalType}}
        child {node {\ttf Ranged/Volume Quantities}
            child { node [label=right:{sampled distance from PCA to origin [$\si\cm$]}] {\ttf impactParameter}}
            child { node [label=right:{column depth considered for interaction. [$\si{\g\per\square\cm}$]}] {\ttf totalColumnDepth}}
            child [missing] {}
            child { node [label=right:{distance above the xy-plane [$\si\cm$]}] {\ttf z}}
            child { node [label=right:{distance from the z-axis [$\si\cm$]}] {\ttf radius}}
        }
    }
    child [missing] {}				
    child [missing] {}				
    child [missing] {}
    child [missing] {}			
    child [missing] {}	
    child [missing] {}
    child [missing] {}
    child [missing] {}
    child [missing] {}
    child [missing] {}
    child [missing] {}
    child [missing] {}	
    child [missing] {}
    child [missing] {}
    child [missing] {}
    child { node [label=right:{Event flux-less weight. [$\si{\GeV\square\cm\steradian}$]}] {\ttf LeptonInjectorWeight} } 
    ;
\end{tikzpicture}
\caption{Lepton Injector Monte Carlo event structure.}\label{fig:tiks_struct}
\end{figure*}
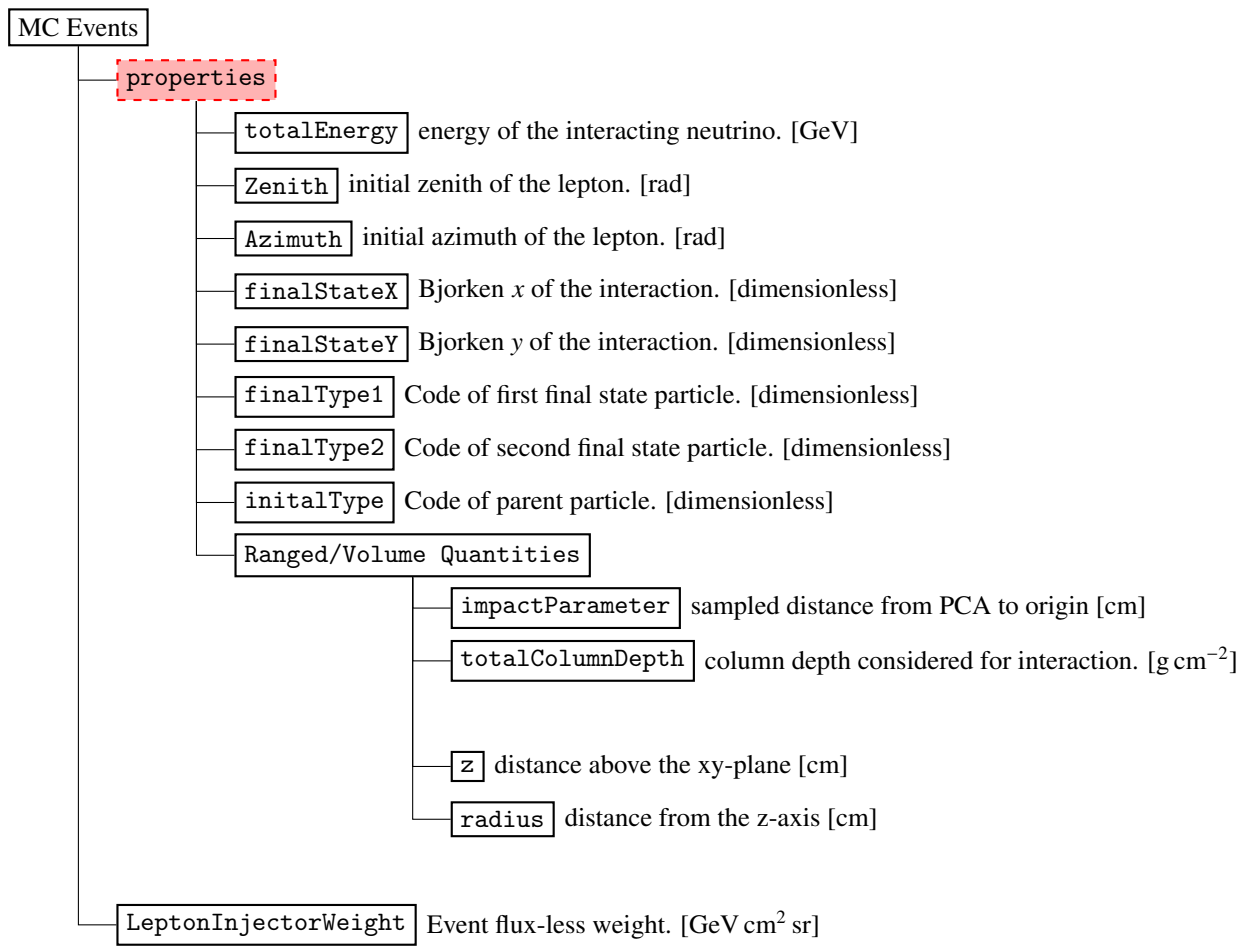

All \LeptonInjector{} events from a single process are saved to a single \hdf~file. Each \textbf{Injector} used in the generation process is given its own dataset inside the \hdf~file with four lists containing an entry for each event generated. Two lists are stored for the two final state particles' parameters, a third list contains the initial state particles' parameters, and a fourth list contains overall parameters for the events.
Each of the event-entries in each of the three lists of particles contain, in order: a number differentiating whether it is initial or final state, the particles' PDG ID~\cite{PhysRevD.98.030001}, the particles' positions, the particles' directions in radians, and the particles' energies. 
The overall properties stored for each event are shown in Figure~\ref{fig:tiks_struct}.
The impact parameter and total column depth are defined in Section~\ref{sec:injection} and shown graphically in Figures~\ref{fig:lepton_ranged2} and~\ref{fig:lepton_ranged4}.

\section{LIC File Structure~\label{sec:lic_structure}}

Data serialized in the LIC file is written little-endian, regardless of machine architecture. 
When a LIC file is first opened, the \textbf{Controller} either overwrites any existing file with the same destination name or begins appending to the end of such an existing file. 
This behavior follows according to user-specification. 
If a new file is being written or an existing one overwritten, a block is first written to the file enumerating all \LeptonInjector{} particle types. 
A header is first written specifying the size of the block, the name of the enumeration, and the length of the enumeration. 
Then the name and number of each entry in the particle enumeration is written. 

Afterwards, each time a new \textbf{Generator} is prepared, the \textbf{Controller} writes a new block to the LIC file. 
Each of these blocks is prefaced with a header specifying the size of the block, the name of the block, and the version of the \LeptonInjector{} serialization code used to write the block. 
Then, all relevant generation parameters are written to the block. 
\section{Earth Model Density\label{sec:earthdensity}}

\begin{figure}[th!]
    \centering
    \includegraphics[width=0.8\linewidth]{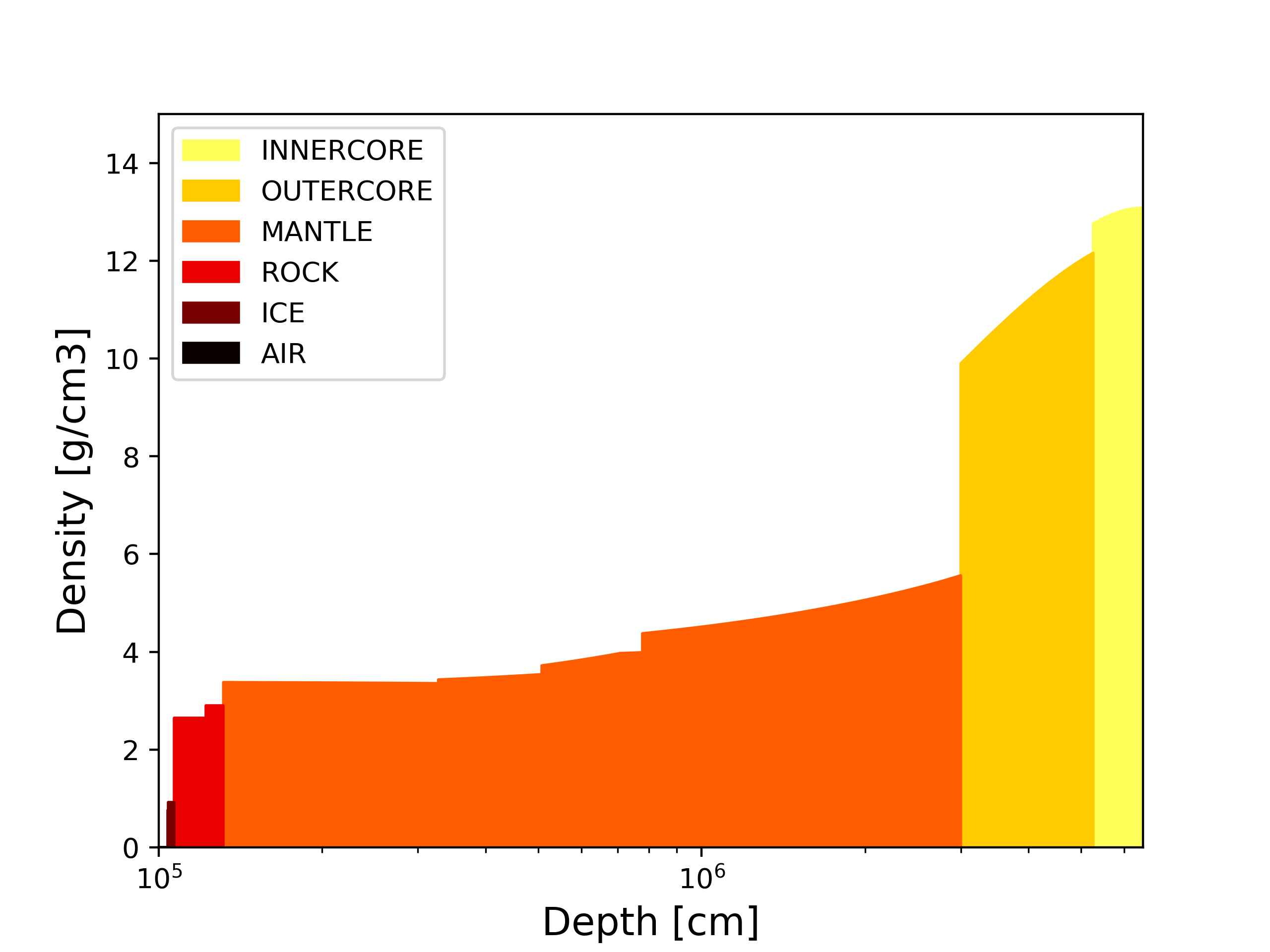}
    \caption{The density of the \LeptonInjector{} Earth model as a function of depth from the edge of Earth's atmosphere.}
    \label{fig:earth_density}
\end{figure}

\LeptonInjector{} uses a modified Preliminary Reference Earth Model (PREM) for column depth calculations around the injection region. The density profile of which is shown in Figure~\ref{fig:earth_density}.

\section{Weighting}\label{sec:weight_append}

The generation procedure produces a set of neutrino properties that include the position, direction, energy, neutrino type, interaction type, and final state kinematic properties.
The distribution of these properties at generation may differ from those we would expect in a physical scenario, so the weighting procedure is designed to correct for these differences.
Beyond the distribution differences, weighting also corrects for differences in dimensionality and raw numbers of events.
In our prototypical scenario, the weight of an event is dimensionless so only a correction factor to the total number of events is needed ($N_\textrm{physical}/N_\textrm{gen}$).

In the case of ranged injection we can separate the generation probability density of an event into several independent components
\begin{equation}
p_\textrm{gen} = p_\textrm{gen}^\textrm{neutrino type} \times p_\textrm{gen}^\textrm{interaction type} \times p_\textrm{gen}^\textrm{energy} \times p_\textrm{gen}^\textrm{direction} \times p_\textrm{gen}^\textrm{impact} \times p_\textrm{gen}^\textrm{depth} \times p_\textrm{gen}^\textrm{kinematics}.
\end{equation}
The generators in \LeptonInjector{} only deal with one neutrino and interaction type each, so $p_\textrm{gen}^\textrm{neutrino type}$ and $p_\textrm{gen}^\textrm{interaction type}$ will either be one or zero depending on if the event matches what can be produced by the generator.
Similarly, $p_\textrm{gen}^\textrm{energy}$ is the probability distribution of injected neutrino energies which is zero for events with neutrino energies outside the bounds of the generator.
Directions are distributed uniformly in ranged injection and so $p_\textrm{gen}^\textrm{direction} = 1/\Omega_\textrm{gen}$ where $\Omega_\textrm{gen}$ is the total solid angle available to the generator.
The $p_\textrm{gen}^\textrm{impact}$ term is the probability distribution related to the impact parameter and angle; since events are sampled uniformly on a disk, this term is the inverse of the disk area, $p_\textrm{gen}^\textrm{impact} = 1/A_\textrm{gen}$, for events intersecting the disk and zero otherwise.
Events are sampled uniformly with respect to column depth along the considered line segment, so the positional distribution of events can be described as $p_\textrm{gen}^\textrm{depth} = \rho_\textrm{gen}(\ell)/X_\textrm{gen}^\textrm{col}$, where $\rho_\textrm{gen}(\ell)$ is the local target mass density where the event is injected and $X_\textrm{gen}^\textrm{col}$ is the total column depth of targets along the considered line segment.
Finally, $p_\textrm{gen}^\textrm{kinematics}$ is the probability distribution of the events kinematic variables; for charged current and neutral current events this is $p_\textrm{gen}^\textrm{kinematics} = (\partial_{xy}\sigma_\textrm{gen}^{\textrm{tot}, i}) / (\sigma_\textrm{gen}^{\textrm{tot}, i})$.

These terms in the generation probability must then be paired with their physical counterparts.
Since our hypothesis can specify the number of neutrinos per type, the neutrino type can be neglected beyond this number correction and the $p_\textrm{gen}^\textrm{neutrino type}$ term from the generator.
The energy, direction, and impact terms all have their counterpart in the neutrino flux $\Phi_\textrm{physical}$, which specifies the physical neutrino distribution in energy, direction, area, and time.
The flux, when paired with a detector livetime $L_\textrm{physical}$ also specifies the total number of neutrinos $N_\textrm{physical}$ by the relation
\begin{equation}
L_\textrm{physical} \times \Phi_\textrm{physical} = N_\textrm{physical} \times p_\textrm{physical}^\textrm{neutrino type} \times p_\textrm{physical}^\textrm{energy} \times p_\textrm{physical}^\textrm{direction} \times p_\textrm{physical}^\textrm{impact}.
\end{equation}

The remaining terms, $p_\textrm{gen}^\textrm{interaction type}$, $p_\textrm{gen}^\textrm{depth}$, and $p_\textrm{gen}^\textrm{kinematics}$, deal with the neutrino interaction itself, which requires special care.
The generation process assumes that the neutrino interacts within a certain region and with a specific interaction type.
In reality, neutrinos on a path to the detector are potentially subject to any of several different interactions, and may pass through the Earth entirely unimpeded.
Thus, we need to account for the probability that the neutrino in the physical scenario would interact within the region considered by the generator $p_\textrm{physical}^\textrm{interaction}$, the depth distribution of all neutrino interactions within that region $p_\textrm{physical}^\textrm{depth}$, the probability that a specific interaction occurs once the interaction point has been chosen $p_\textrm{physical}^\textrm{interaction type}$, and finally the kinematic distribution $p_\textrm{physical}^\textrm{kinematics}$.
The former two terms, $p_\textrm{physical}^\textrm{interaction}$ and $p_\textrm{physical}^\textrm{depth}$ depend explicitly on the line segment considered by the generator when choosing the neutrino interaction vertex.
The interaction probability can be cast in terms of the ``survival'' probability, $p_\textrm{physical}^\textrm{interaction} = 1-p_\textrm{physical}^\textrm{survival}$, the probability that the neutrino will pass through the region without interacting.
The survival probability is given by
\begin{equation}
p_\textrm{physical}^\textrm{survival} = \exp\left(-{\int_{\ell_i}^{\ell_f}{d\ell \sum_{p,i} n_\textrm{physical}^p(\ell) \sigma_\textrm{physical}^{\textrm{tot},p,i} }}\right),
\end{equation}
where $p$ iterates over the possible targets (usually nucleons and electrons), $i$ iterates over interaction types, and $n_\textrm{physical}^p(\ell)$ is the density of target $p$ at a point $\ell$ along the considered line segment
Thus,
\begin{equation}
p_\textrm{physical}^\textrm{interaction} = 1-\exp\left(-{\int_{\ell_i}^{\ell_f}{d\ell \sum_{p,i} n_\textrm{physical}^p(\ell) \sigma_\textrm{physical}^{\textrm{tot},p,i} }}\right).
\end{equation}
Another way of writing this is in terms of the total column depth for each target $X_\textrm{physical}^{\textrm{col}, p}$, target molar mass $M_p$, and Avagadro's number $N_A$, such that
\begin{equation}
p_\textrm{physical}^\textrm{interaction} = 1-\exp\left(-N_A{\sum_{p,i} (X_\textrm{physical}^{\textrm{col}, p}/M_p) \sigma_\textrm{physical}^{\textrm{tot},p,i} }\right).
\end{equation}
The implementation in \LeptonWeighter{} groups protons and neutrons together and assumes that the molar mass of nucleons is $\SI{1}{\gram\per\mol}$.
The depth distribution $p_\textrm{physical}^\textrm{depth}$ follows a similar form as the survival probability, but is normalized within the generation bounds such that
\begin{equation}
p_\textrm{physical}^\textrm{depth} = \exp\left(-{\int_{\ell_i}^{\ell}{d\ell \sum_{p,i} n_\textrm{physical}^p(\ell) \sigma_\textrm{physical}^{\textrm{tot},p,i} }}\right) / \int_{\ell_i}^{\ell_f}{d\ell}\exp\left(-{\int_{\ell_i}^{\ell}{d\ell \sum_{p,i} n_\textrm{physical}^p(\ell) \sigma_\textrm{physical}^{\textrm{tot},p,i} }}\right),
\end{equation}
which can similarly be recast in terms of the density or column depth.
The last two terms, $p_\textrm{physical}^\textrm{interaction type}$ and $p_\textrm{physical}^\textrm{kinematics}$, depend only on the position and type of the interaction, and so are independent of the generator.
Once we assume that an interaction occurs at a known location, the probability of a specific interaction $p_\textrm{physical}^\textrm{interaction type}$ is the ratio of total cross sections at that location $p_\textrm{physical}^\textrm{interaction type} = (\sigma_\textrm{physical}^{\textrm{tot}, i})/(\sum_j \sigma_\textrm{physical}^{\textrm{tot}, j})$ where the subscript $j$ iterates over all possible interactions.
For the chosen interaction type, the kinematic distribution is simply $p_\textrm{physical}^\textrm{kinematics} = (\partial_{xy}\sigma_\textrm{physical}^{\textrm{tot}, i}) / (\sigma_\textrm{physical}^{\textrm{tot}, i})$ in the case of charged or neutral current interactions.

By pairing up the terms we can see all the effects that are accounted for in the event weight
\begin{equation}
w_\textrm{MC} = \frac{N_\textrm{physical}}{N_\textrm{gen}} p_\textrm{physical}^\textrm{interaction} \frac{p_\textrm{physical}^\textrm{neutrino type}}{p_\textrm{gen}^\textrm{neutrino type}} \frac{p_\textrm{physical}^\textrm{interaction type}}{p_\textrm{gen}^\textrm{interaction type}} \frac{p_\textrm{physical}^\textrm{energy}}{p_\textrm{gen}^\textrm{energy}} \frac{p_\textrm{physical}^\textrm{direction}}{p_\textrm{gen}^\textrm{direction}} \frac{p_\textrm{physical}^\textrm{impact}}{p_\textrm{gen}^\textrm{impact}} \frac{p_\textrm{physical}^\textrm{depth}}{p_\textrm{gen}^\textrm{depth}} \frac{p_\textrm{physical}^\textrm{kinematics}}{p_\textrm{gen}^\textrm{kinematics}}.
\end{equation}
Practical implementations of this replace some of the physical terms with the flux and livetime to obtain the event weight
\begin{equation}
w_\textrm{MC} = \frac{L_\textrm{physical}\Phi_\textrm{physical}}{N_\textrm{gen} p_\textrm{gen}^\textrm{neutrino type} p_\textrm{gen}^\textrm{energy} p_\textrm{gen}^\textrm{direction} p_\textrm{gen}^\textrm{impact}} \times p_\textrm{physical}^\textrm{interaction} \times \frac{p_\textrm{physical}^\textrm{interaction type}}{p_\textrm{gen}^\textrm{interaction type}} \times \frac{p_\textrm{physical}^\textrm{depth}}{p_\textrm{gen}^\textrm{depth}} \times \frac{p_\textrm{physical}^\textrm{kinematics}}{p_\textrm{gen}^\textrm{kinematics}}.
\end{equation}

When simulation is created using multiple generators we must consider the probability that a particular event may be generated in either generator, regardless of which generator it originated from.
The behavior we desire is such that events from non-overlapping regions of the parameter space retain their original single-generator weights, but events that reside in the overlap regions are down-weighted to account for the overlap.
The event weight then takes the form
\begin{equation}
w_\textmd{MC} = \left[\sum_i \left(p_\textmd{physical}^{i}\right)^{-1} \times p_\textmd{gen}^{i}\right]^{-1},
\end{equation}
where the superscript $i$ denotes the different generators, $p_\textmd{physical}^{i}$ is the physical contribution to the weighting, and $p_\textmd{gen}^{i}$ is the generation contribution to the weighting.
Note the superscript $i$ on the physical contribution is there because the terms $p_\textrm{physical}^\textrm{interaction}$ and $p_\textrm{physical}^\textrm{depth}$ depend on the particular generator.
If the generation settings governing the line segment along which the event is injected are common to all generators then this dependence can be dropped and $p_\textrm{physical}$ can be factored out.

The above description in the weighting starts from the ranged injection procedure, but only minor modifications are needed for this to be applicable to volume injection.
The difference arises from the $p_\textrm{gen}^\textrm{depth}$ and $p_\textrm{gen}^\textrm{impact}$ terms on the generation side and $p_\textrm{physical}^\textrm{depth}$ and $p_\textrm{physical}^\textrm{interaction}$ on the physical side.
These generation terms are directly analogous to steps in the ranged injection procedure where the position of closest approach and interaction vertex position are chosen.
In the volume injection, the interaction vertex is chosen in a single step, so we can replace these two generation terms with the single $p_\textrm{gen}^\textrm{position}$ term which is a uniform probability density along the line segment within the injection cylinder and zero outside.
The physical terms only differ in that the line segment considered for the calculation now must come from the volume injection procedure.
Specifically, the line segment considered passes through the interaction vertex following the injected neutrino direction, beginning and ending at the boundaries of the injection cylinder.

These differences between the ranged and volume injection mean that the physical contributions to the weighting will differ between the two, and must be calculated separately for each event if both methods are used in the event generation.

Finally, the approximation used in Eq.~\ref{eq:approx_weight} can be obtained by expanding the depth and interaction terms for a vanishing product of the interaction cross section and column depth.
For the column depths and cross sections used in \LeptonInjector{}, this approximation remains valid for sub $\si\ZeV$ neutrino energies.

\section{Generation Example\label{sec:example_generation}}

This \Python example is included in the \LeptonInjector{} source code.
It creates an Injector in Ranged mode for producing CC muon-neutrino events of initial energy between $\SI{1000}\GeV$ to $\SI{100000}\GeV$, and outputs the data to an \hdf~file. 
\pythonexternal{examples/inject_example.py}

\section{Weighting Example\label{sec:example_weighting}}

The following \Python example is included in \LeptonWeighter{}.
It reads a set of generated events and computes the weights of for a given neutrino cross sections and fluxes.
The result is stored in an HDF5 file for later usage.
\pythonexternal{examples/weighting_example_python.py}

\end{document}